\documentclass[12pt]{iopart}
\usepackage{graphicx}
\usepackage{support-caption}
\usepackage[labelfont=bf, textfont=it]{caption}
\usepackage[labelfont=bf, textfont=it]{subcaption}
\usepackage[backend=biber, style=numeric-comp, sorting=none, maxcitenames=2, uniquelist=false]{biblatex}
\usepackage{hyperref}
\usepackage{amssymb}

\usepackage{perpage} %the perpage package
\MakePerPage{footnote} %the perpage package command

\bibliography{ETG_MASTU_2}

\begin{document}

\title[Sensitivities of electron-scale core transport in MAST Upgrade]{Sensitivities of electron-scale core transport in MAST Upgrade}

\author{B.S. Patel$^{1}$\footnote{Present address: Gridfire Inc, 3422 Old Capitol Trail, Suite 700, New Castle, Wilmington
19808-6124 Delaware, USA}, T. Adkins$^{2}$, S. Blackmore$^{1}$, D. Kennedy$^{1}$, C. Vincent$^{1}$}
%\email{bhav.patel@ukaea.uk}
\address{$^1$ UKAEA (United Kingdom Atomic Energy Authority), Culham Campus, Abingdon, Oxfordshire, OX14 3DB, United Kingdom}
\address{$^2$ Princeton Plasma Physics Laboratory, Princeton University, Princeton, NJ 08540, United States of America}

%
%\author{B. Patel}

%\address{York Plasma Institute, University of York, YO10 5DD}
\ead{bhavin@gridfire.co}

\vspace{10pt}
\begin{indented}
\item[]August 2026
\end{indented}

\begin{abstract}

Integrated modelling of the Mega Ampere Spherical Tokamak Upgrade (MAST-U) indicates that turbulent electron heat transport can dominate over ion transport across a range of operating regimes. Only a limited set of instabilities are able to produce this behaviour, with the primary candidates being microtearing modes (MTMs) and electron temperature gradient (ETG) driven modes. This work investigates electron-scale core transport in two L-mode plasmas and one H-mode plasma on MAST-U using local gyrokinetic analysis. The linear and nonlinear sensitivities of ETG modes are examined, with particular focus on their dependence on electron temperature gradients and $\mathrm{E \times B}$ shearing rates. In L-mode discharges, ETG modes are found to be linearly unstable over a broad radial region and can drive experimentally relevant levels of electron heat transport, particularly towards the outer core. The transport is strongly sensitive to both the electron temperature gradient and the $\mathrm{E \times B}$ shear, with nonlinear simulations showing stiff transport and good agreement with experimental estimates within uncertainty. Towards the core, however, ETG-driven transport is reduced and is insufficient to fully explain the observed anomalous heat flux. In contrast, in the H-mode plasma ETG modes are found to be stable or only weakly unstable in the core, resulting in negligible electron-scale transport. This is attributed to higher equilibrium pressure gradients and reduced electron temperature gradients, both of which act to stabilise ETG turbulence. Following the collapse of the core rotation profile, ETG modes can become unstable towards the outer core, but typically remain insufficient to account for the full level of transport. Overall, these results show that ETG turbulence can play a significant role in setting electron heat transport in L-mode plasmas, particularly at larger radii, but is unlikely to dominate in high-performance H-mode conditions.

\end{abstract}

\section{Introduction} \label{sec:intro}

Turbulent transport can be the limiting factor in the performance of spherical tokamaks, and understanding the trends observed in experiments can help optimise the confinement in both existing devices and future reactors \cite{kaye2021thermal}. Integrated modelling of MAST-U using TRANSP \cite{pankin2025transp} shows that many regimes exist in which anomalous electron heat transport dominates over anomalous ion transport, which typically remains close to neoclassical levels. The relatively weak ion transport can be partially attributed to strong rotation shear driven by neutral beam injection (NBI) \cite{roach2009gyrokinetic, kaye2021thermal}. 

Only turbulence driven by certain underlying linear instabilities can produce transport signatures consistent with this behaviour, with the primary candidates being electron temperature gradient (ETG) modes and larger-scale microtearing modes (MTMs). Both have been identified in simulations of spherical tokamaks and are capable of driving strong electron heat transport while leaving ion transport comparatively low \cite{colyer2017collisionality, roach2009gyrokinetic, giacomin2023nonlinear, patel2025impact, applegate2004microstability, guttenfelder2012scaling}.

This work focuses on electron-scale turbulence, specifically ETG modes, while ion-scale instabilities will be addressed in a follow-up study \cite{kennedy2026tbd}. Using the gyrokinetic code CGYRO \cite{candy2016high}, we examine the linear stability and nonlinear transport driven by ETG modes in the core of MAST-U plasmas. The analysis considers two low-power L-mode discharges and one high-power H-mode discharge. In the L-mode regime, regions where ETG-driven transport may dominate are first identified using integrated modelling by examining the ratio of anomalous electron to ion heat flux. Linear simulations are then used to characterise the stability and parametric sensitivities of ETG modes, followed by nonlinear simulations to assess their contribution to the experimentally inferred heat flux. In several cases, good agreement with experiment is obtained. The same approach is then applied to the H-mode plasma. However, in this regime ETG modes are generally found to be linearly stable or only weakly unstable, resulting in predicted transport levels that are significantly below experimental estimates. This behaviour is linked to higher equilibrium pressure gradients and reduced electron temperature gradients, both of which act to stabilise ETG turbulence.

The remainder of this paper is structured as follows. Section \ref{sec:l_mode} examines ETG-driven transport in the L-mode discharges, including both linear and nonlinear analysis. Section \ref{sec:h_mode} extends this approach to the H-mode plasma. Conclusions are presented in Section \ref{sec:conclusion}.

\section{L-mode discharges}
\label{sec:l_mode}

\subsection{Nature of transport in L-mode discharges}

\begin{figure}[!htb]
    \begin{subfigure}{0.49\textwidth}
        \centering
        \includegraphics[width=75mm]{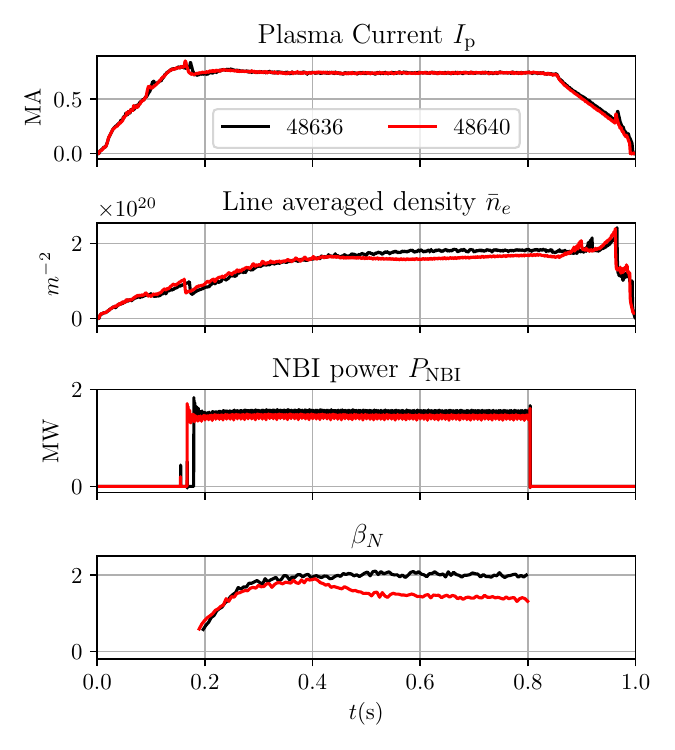}
        \caption{}
        \label{fig:lmode_pulse}
    \end{subfigure}
    \begin{subfigure}{0.49\textwidth}
        \centering
        \includegraphics[width=75mm]{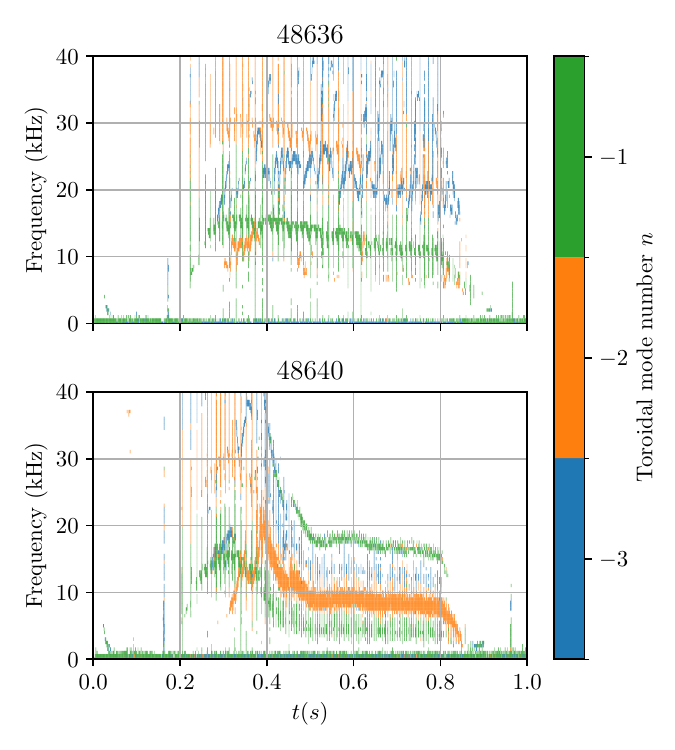}
        \caption{}
        \label{fig:lmode_mhd}
    \end{subfigure}
    \caption{(a) Time traces of the plasma current, line-averaged density, NBI power and $\beta_N$ for two MAST-U discharges \#48636 (black) and \#48640 (red). (b) Toroidal mode number analysis of OMAHA coils on MAST-U illustrating the difference in mode number of MHD fluctuations for \#48636 (top) and \#48640 (bottom).}
    \label{fig:lmode_summary}
\end{figure}

Two MAST-U L-mode discharges, \#48636 and \#48640, are compared, both with $I_\mathrm{p} = 750\,\mathrm{kA}$ and on-axis NBI heating. Figure \ref{fig:lmode_pulse} shows time traces of the plasma current $I_\mathrm{p}$, line-averaged density $\bar{n}_e$, NBI power $P_{\mathrm{NBI}}$, and $\beta_N$. The discharges have similar setups, with \#48640 having slightly lower input power. However, their evolution in $\beta$ differs significantly, which can be attributed to differences in MHD stability. The full profiles of electron density, electron temperature and ion temperature are included for \#48640 in Figure \ref{fig:lmode_profiles} for completeness.

\begin{figure}[!htb]
    \centering
    \includegraphics[width=150mm]{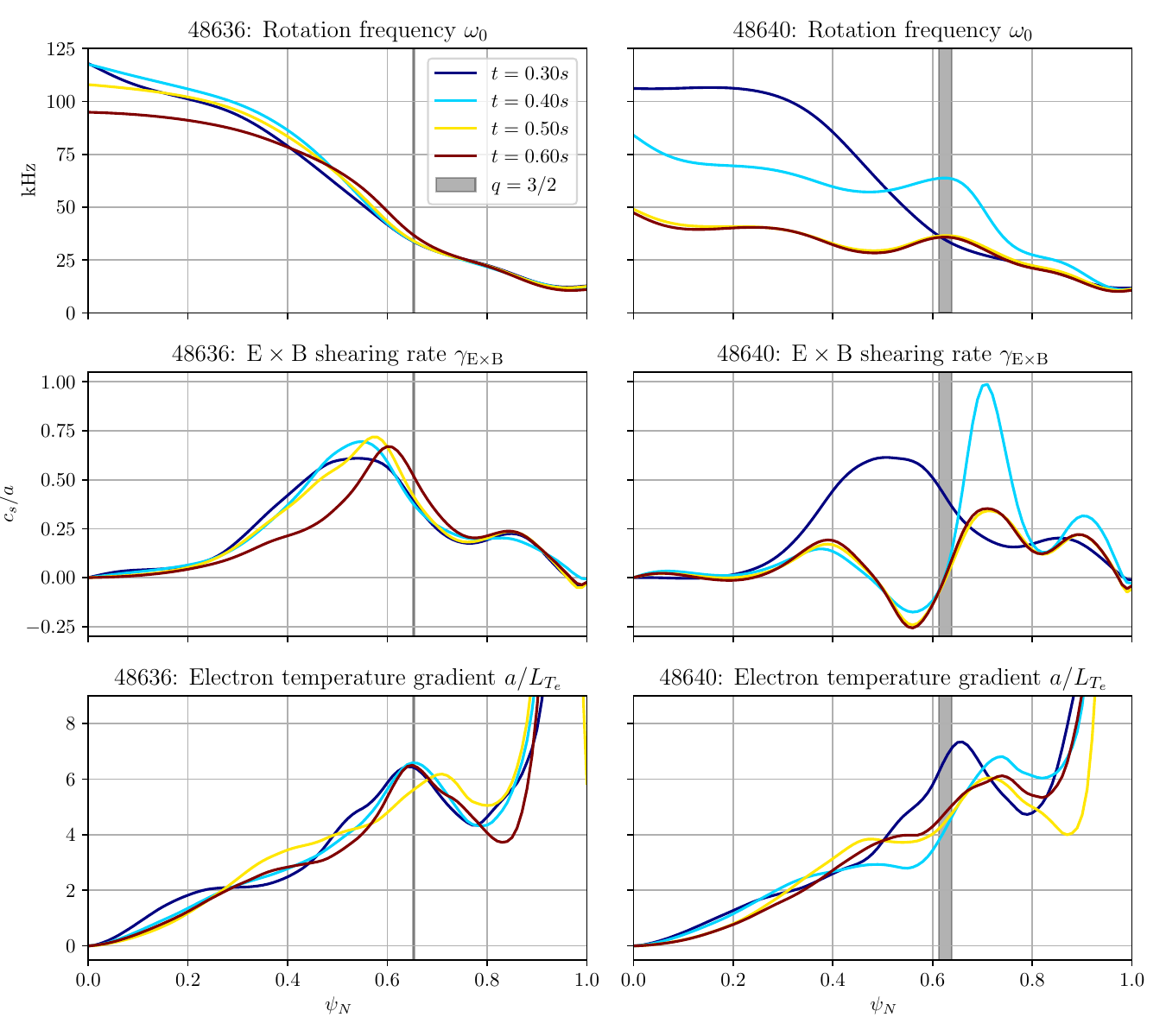}
    \caption{Rotation frequency (top row), $\mathrm{E \times B}$ shearing rate (middle row) and electron temperature gradient (bottom row) at various times in the pulse, denoted by the line colour for \#48636 (left) and \#48640 (right).}
    \label{fig:lmode_rotation}
\end{figure}

Figure \ref{fig:lmode_mhd} shows the toroidal mode number $n$ of the MHD activity measured using the OMAHA coils. In the lower-performing discharge, \#48640, a strong $m/n = 3/2$ neoclassical tearing mode forms at $t \approx 0.3\,\mathrm{s}$. This mode initially spins up but eventually leads to rotational braking of the plasma through neoclassical toroidal viscosity \cite{nowak2012evidence}, effectively pinning the core rotation to the value at the $q = 3/2$ surface. This behaviour is illustrated in Figure \ref{fig:lmode_rotation}, where the top row shows the plasma rotation frequency measured by charge exchange spectroscopy \cite{conway2006high}. In \#48636, the $3/2$ mode disappears by $t=0.4\,\mathrm{s}$ and has little impact on the rotation, allowing the plasma to maintain higher performance. In contrast, \#48640 has much smaller, more frequent sawteeth but exhibits a strong reduction in rotation. The corresponding $\mathrm{E \times B}$ shearing rate, 

\[\gamma_\mathrm{E \times B} = -\frac{r}{q}\frac{\partial \omega_0}{\partial r} 
\]

is shown in the second row of Figure \ref{fig:lmode_rotation}. After the mode pins the rotation profile in \#48640, $\gamma_\mathrm{E \times B}$ is significantly reduced for $\psi_N < 0.65$, while increasing at larger radii due to the outward flux of momentum. $\mathrm{E \times B}$ shear can act to suppress ITG turbulence and ETG turbulence when weakly driven \cite{kaye2021thermal, smith2009observations}, though when ETG is more strongly driven it can be more resilient \cite{roach2009gyrokinetic}. This radial variation in shear is expected to influence electron-scale turbulence by shearing the ion-scale radial streamers that can form via ETG turbulence \cite{jenko2002prediction, belli2024flow}. The electron temperature gradient, $a/L_{T_e}$, is shown in the third row. Relative increases in $\gamma_\mathrm{E \times B}$ correlate with increases in $a/L_{T_e}$.

\begin{figure}[!htb]
        \centering
        \includegraphics[width=75mm]{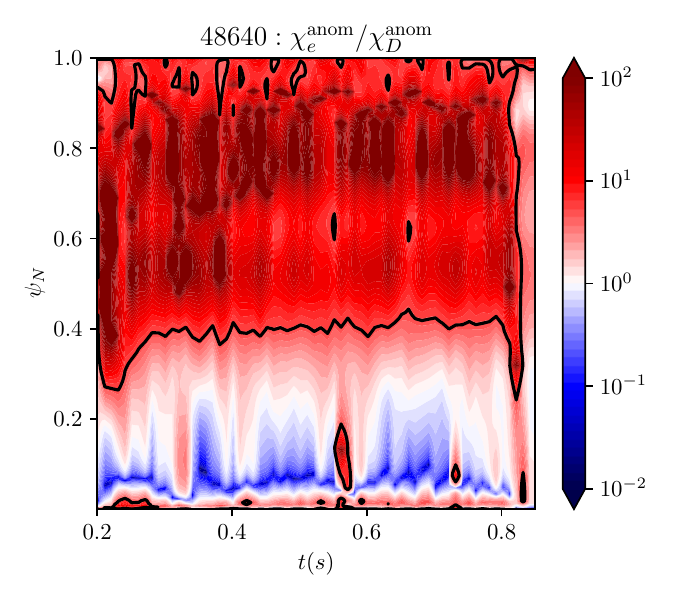}
        \caption{Ratio of electron to deuterium heat diffusivity for \#48640, the pulse with the reduced performance, calculated by TRANSP as a function of time and normalised poloidal flux. The black line denotes $\chi_e^\mathrm{anom}/\chi_D^\mathrm{anom} = 5$.}
        \label{fig:48640_chi_ratio}
\end{figure}

Integrated modelling using TRANSP allows the anomalous transport levels in each species to be estimated. Figure \ref{fig:48640_chi_ratio} shows the ratio of anomalous electron to deuterium heat diffusivities, $\chi_e^{\mathrm{anom}} / \chi_D^{\mathrm{anom}}$, for \#48640. For $\psi_N > 0.3$, the electron diffusivity exceeds that of deuterium, which remains close to neoclassical levels. This behaviour is consistent with ETG modes contributing significantly to transport in this region. MTMs also have a matching transport signature but tend to occur at larger ion scales, so are outside the scope of this work, though they will be the subject of future investigations. At $\psi_N \approx 0.6$, a slight increase in the ion anomalous diffusivity is observed after $t \approx 0.4\,\mathrm{s}$, corresponding approximately to the location of the $q = 3/2$ surface where $\gamma_\mathrm{E \times B} \sim 0$. In the deep core, however, the anomalous deuterium transport dominates over the anomalous electron transport, suggesting a different regime of turbulence.

Given the presence of strong MHD activity, it is challenging to fully disentangle transport arising from MHD modes and that from turbulence. Therefore, the following analysis focuses primarily on regions away from the tearing mode location ($\psi_N\approx0.6$, $q=3/2$), where gyrokinetic modelling is more likely to be valid.

\subsection{Numerical details}
Simulations were initially performed fully electromagnetically, including three fluctuating fields: $\phi$, $A_{||}$, and $B_{||}$, although certain fields are later shown to be negligible. Four kinetic species were included: electrons, deuterium, carbon, and a `fast' deuterium population generated by NBI. As CGYRO supports only Maxwellian species, this fast population was modelled as a hot thermal species with equivalent energy content.

Linear simulations were performed using 64 grid points along the field line, with 12 connected $2\pi$ segments, 24 pitch angles, and 8 energy grid points. All analysis of the gyrokinetic simulations was carried out using Pyrokinetics \cite{patel2024pyrokinetics}. The default Pyrokinetics conventions were adopted, where the binormal wavenumber is defined as $k_y = (n /B_0) (\partial \psi /\partial r)^{-1}$, and the ion Larmor radius is given by $\rho_s = c_s/(eB_0 /m_D)$, with $n$ being the toroidal mode number, $B_0 = f(\psi)/R_\mathrm{major}$ and the ion sound speed $c_s = \sqrt{T_e/m_D}$. Here, $f(\psi)$ is the poloidal current flux function in the Grad--Shafranov equation and the normalising length scale is the minor radius $a$.

\subsection{Linear stability in L-mode}
In this section, the stability of the L-mode discharge, \#48640, is examined both before and after the core rotation collapse, at $t = 0.3\,\mathrm{s}$ and $t = 0.6\,\mathrm{s}$, respectively. The most relevant input parameters can be found in Tables \ref{tab:48640_t03} and \ref{tab:48640_t06}, with the full input files available in \ref{app:a}.

\begin{figure}[htb]
    \centering
    \includegraphics[width=150mm]{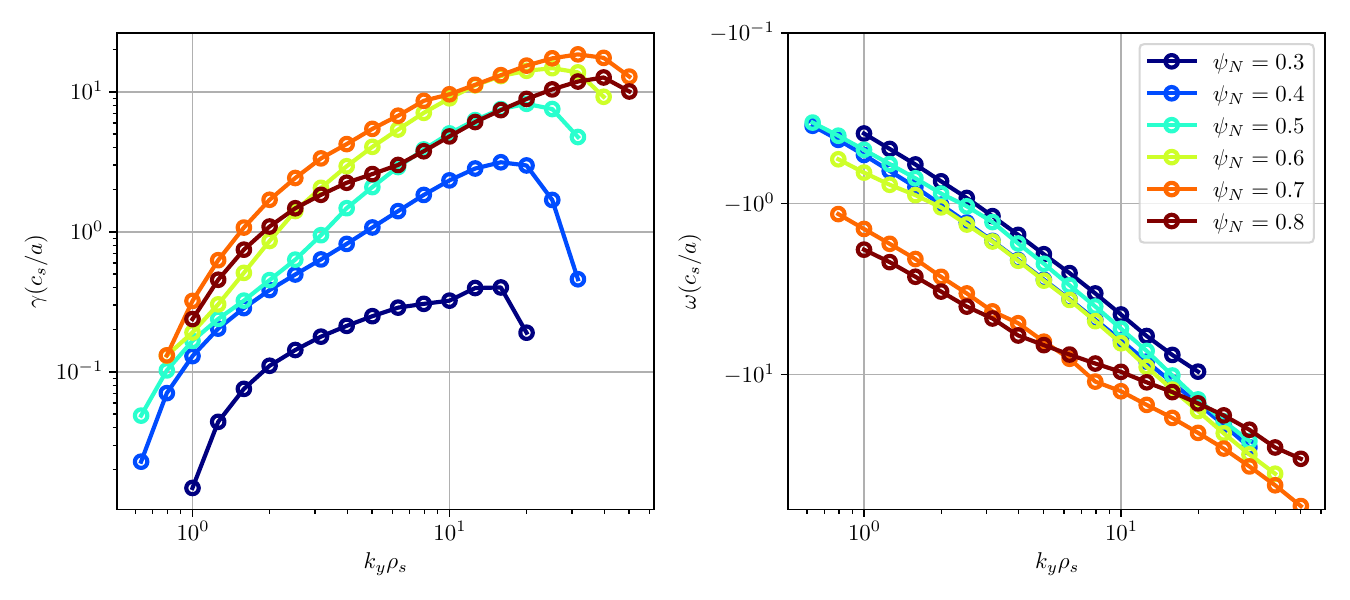}
    \caption{Growth rate (left) and mode frequency (right) of ETG modes in \#48640 at $t = 0.3\,\mathrm{s}$. The colour indicates the flux surface of the simulations. A clear stabilisation is observed in the deep core where the electron scale is completely stable at $\psi_N<0.3$, with peak destabilisation towards the edge around $\psi_N=0.7$.}
    \label{fig:lmode_all_surfaces}
\end{figure}

Figure \ref{fig:lmode_all_surfaces} shows the linear stability at electron scales ($k_y \rho_s > 0.5$) across a range of flux surfaces in the core plasma at $t = 0.3\,\mathrm{s}$. Moving towards the core, the ETG modes become increasingly stabilised and are completely stable for $\psi_N < 0.3$. The growth rate peaks at $\psi_N = 0.7$. These trends are consistent with the diffusivities predicted by TRANSP, which show a decrease in $\chi_e^\mathrm{anom}/\chi_D^\mathrm{anom}$ towards the core and an increase towards the edge. This behaviour also correlates with the $a/L_{T_e}$ profile shown in Figure \ref{fig:lmode_rotation}.

\begin{figure}[!htb]
    \centering
    \includegraphics[width=75mm]{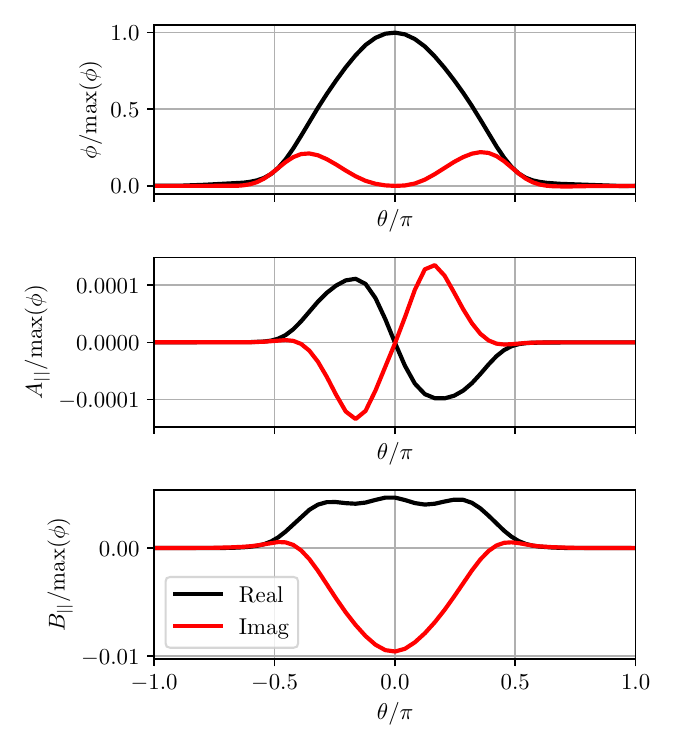}
    \caption{Eigenfunctions of the ETG mode at the peak growth rate at $k_y\rho_s=37$ for $\psi_N=0.75$. Note the simulation domain is extended further along the field line but only the region $-\pi \leq \theta \leq \pi$ is shown as the eigenfunction tails off rapidly.}
    \label{fig:etg_eigfunc}
\end{figure}

Towards the edge of the plasma, at $\psi_N = 0.75$, the eigenfunctions at the peak growth rate ($k_y \rho_s = 37$) are shown in Figure \ref{fig:etg_eigfunc}, with the dimensionless fields being normalised to the peak $\phi$ value. The mode exhibits a narrow eigenfunction, with $\phi$ significantly larger than the electromagnetic components, and is localised around $\theta = 0$ rather than at the large $\hat{s}\theta$ characteristic of the toroidal pedestal modes reported in \cite{parisi2020toroidal}. The parallel extent of the mode narrows monotonically with increasing $k_y\rho_s$, from $\Delta\theta \sim 1.5\pi$ at $k_y\rho_s \approx 2$ to $\Delta\theta \sim 0.5\pi$ at $k_y\rho_s \approx 46$, consistent with increasingly strong finite-Larmor-radius damping of the tail of the eigenfunction as $k_\perp\rho_e$ grows along the field line. The identification of the underlying drive is addressed in Section \ref{sec:lin_sens} through tests in which the electron magnetic drift frequency is artificially removed. Ion-scale modes are found to be unstable for all flux surfaces but will be the focus of future work.

\subsubsection{Linear sensitivities in L-mode}
\label{sec:lin_sens}

\begin{figure}[!htb]
    \centering
    \includegraphics[width=150mm]{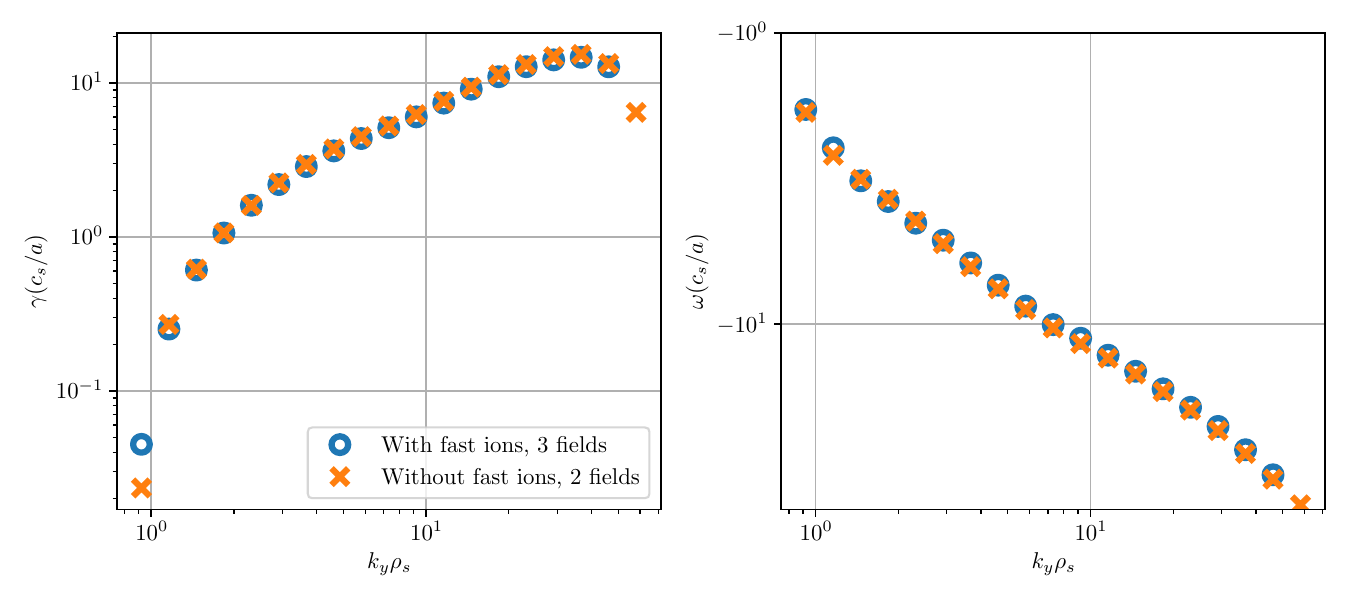}
    \caption{Linear $k_y$ spectrum for $\psi_N=0.75$ with all four kinetic species and three fields (blue circles), compared to a simulation with thermalised fast ions and no $B_{||}$ fluctuations (orange crosses).}
    \label{fig:48640_nofast}
\end{figure}

The sensitivity of ETG modes to electron kinetic gradients is examined on the flux surface at $\psi_N = 0.75$. The linear $k_y$ spectrum is shown in Figure \ref{fig:48640_nofast} for simulations including all four kinetic species, and for a case where fast ions are removed and $B_{||}$ fluctuations are neglected. The spectra are largely unchanged, with only a slight stabilisation observed at low $k_y\rho_s < 1.0$, where fast ion physics is relevant due to the larger scales rather than at smaller scales where the gyro-orbit of the fast ion essentially averages out any small-scale fluctuation. This indicates that fast ions and $B_{||}$ fluctuations have minimal impact on linear stability at electron scales for this equilibrium. Note that although fast ions are removed kinetically, their contribution to the equilibrium pressure gradient,
\[
\beta' = \frac{2\mu_0}{B_0^2} \frac{\partial p}{\partial r},
\]
is retained. Here, $\beta' < 0$ for conventional pressure profiles. In the gyrokinetic equation, the pressure gradient enters both the gradient drive and the drift frequency, and ideally the two should be derived from the same pressure profile, but in practice they are often specified independently. $\beta'$ refers specifically to the pressure gradient used in the drift frequency, while gradients such as $a/L_T$ and $a/L_n$ refer to the terms used in the gradient drive. In this work, $\beta'$ is fixed during kinetic gradient scans and set such that it is consistent with the Grad-Shafranov solution derived from experiment. If instead $\beta'$ is adjusted self-consistently after removing fast ions, a slight destabilisation is observed. However, this effect remains small due to the low $\beta_e$. Therefore, perpendicular magnetic fluctuations and fast ions are neglected in the L-mode simulations to reduce computational cost.

\begin{figure}[!htb]
    \centering
    \includegraphics[width=150mm]{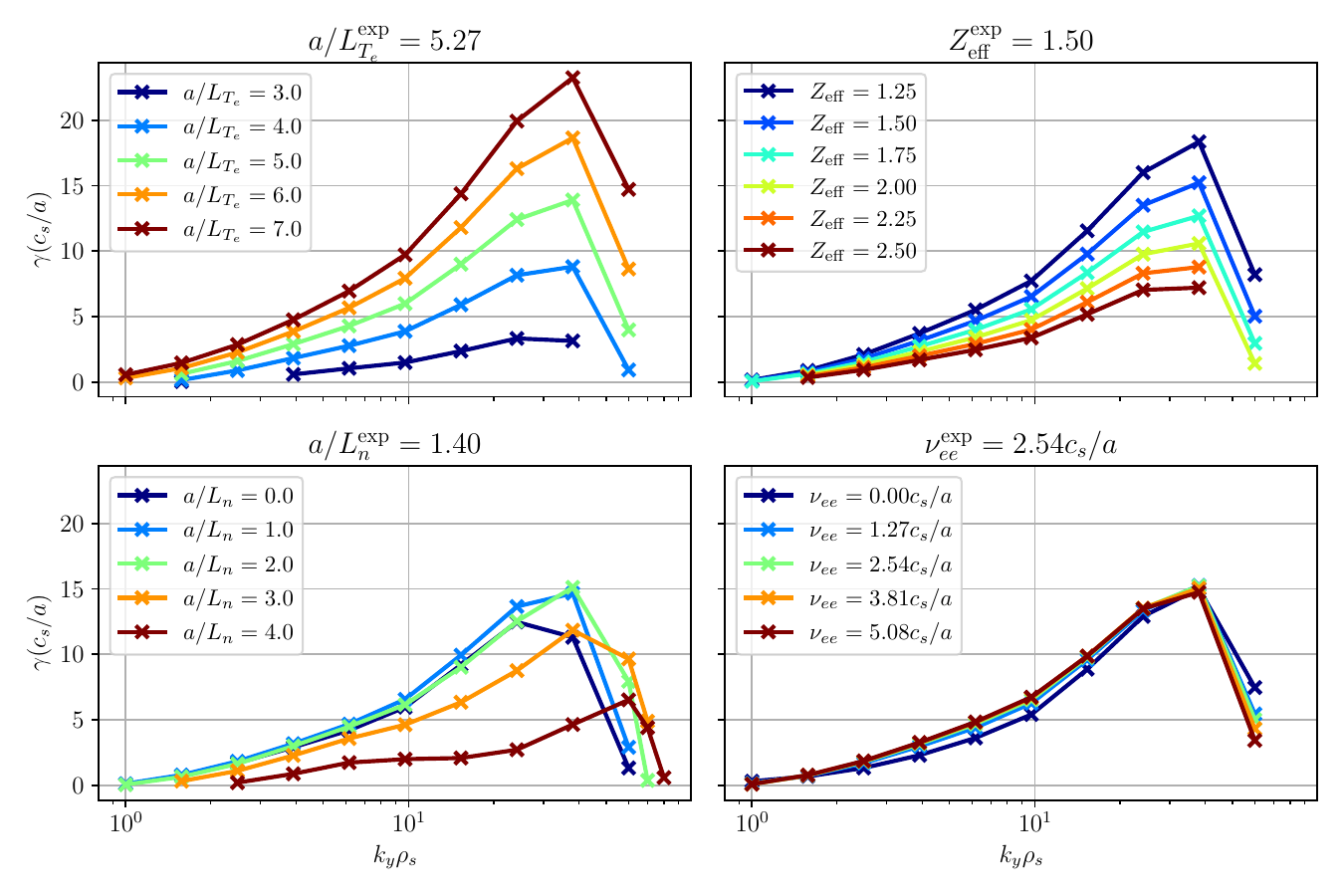}
    \caption{Sensitivity of the ETG mode to $a/L_{T_e}$ (top left), $Z_\mathrm{eff}$ (top right), $a/L_{n_e}$ (bottom left), and $\nu_{ee}$ (bottom right). The ETG modes are most sensitive to $a/L_{T_e}$, which is strongly destabilising, while $Z_\mathrm{eff}$ is found to be strongly stabilising.}
    \label{fig:linear_etg_scans}
\end{figure}

The sensitivity of ETG growth rates to kinetic profile variations is shown in Figure \ref{fig:linear_etg_scans}. To preserve quasi-neutrality, density gradients of all species are varied simultaneously. The ETG modes are strongly destabilised by increasing $a/L_{T_e}$, although the location of the peak growth rate in $k_y\rho_s$ remains unchanged. Note that in this temperature gradient scan, $\eta_e = (a/L_{T_e}) / (a/L_{n_e})$ remains above 1 throughout. In contrast, varying the density gradient $a/L_{n_e}$ has a more nuanced effect. For $\eta_e \gg 1$, the peak growth rate and its location are unaffected. However, as $\eta_e \rightarrow 1$ (corresponding to $a/L_{n_e} \rightarrow 5.27$), the modes are stabilised and the peak shifts to slightly higher $k_y\rho_s$. For $a/L_{n_e} \geq 5.0$, the mode becomes fully stable (not shown). 

Scans of $a/L_{n_e}$ at fixed $\eta_e$ (not shown here) show no shift in the peak location. This behaviour is consistent with previous JET pedestal ETG studies \cite{parisi2020toroidal, chapman2022role}. As $\eta_e$ approaches unity, it determines the location of the peak growth rate, whereas the peak location remains unchanged when $\eta_e$ is well above unity \cite{jenko2001critical}. The present surface lies well above this threshold, at $\eta_e = 3.8$. Artificially removing the electron magnetic drift suppresses the modes reported here (not shown), leaving only a much weaker branch at $k_y\rho_s \approx 46$ with $\gamma \approx 3\,c_s/a$. The dominant instability is therefore a toroidal ETG mode, driven predominantly by the magnetic drifts rather than by parallel streaming, which is instead stabilising for toroidal modes.

Impurities influence ETG turbulence primarily through changes in collisionality and effective charge, $Z_\mathrm{eff}$, and are generally stabilising \cite{reshko2008effects, romanelli2011linear}. As $Z_\mathrm{eff}$ measurements were unavailable for these discharges, a value of $Z_\mathrm{eff} = 1.5$ was assumed in the transport analysis. This introduces significant uncertainty, motivating a sensitivity study. Here, $Z_\mathrm{eff}$ is varied by increasing the carbon concentration, while collisionality is independently adjusted via $\nu_{ee}$\footnote{CGYRO will set all self and inter-species collision frequencies consistently relative to $\nu_{ee}$}. Increasing $Z_\mathrm{eff}$ enhances electrostatic shielding through the adiabatic ion response \cite{reshko2008effects, jenko2001critical, adkins2026asymptotic}, leading to clear stabilisation of the ETG modes (Figure \ref{fig:linear_etg_scans}, upper right). This highlights the importance of accurate $Z_\mathrm{eff}$ measurements for determining critical thresholds.

Collisionality has a comparatively weak effect. A slight destabilisation is observed at $k_y\rho_s < 10$, though the peak growth rate is largely unchanged. At $k_y\rho_s \leq 1$, the trend reverses, with increased collisionality leading to stabilisation. This behaviour is likely linked to hybridisation with ion-scale modes (e.g.\ TEM/UM), where increased collisionality enhances detrapping and reduces the drive \cite{shen2019properties}.

\begin{figure}[!htb]
    \centering
    \includegraphics[width=150mm]{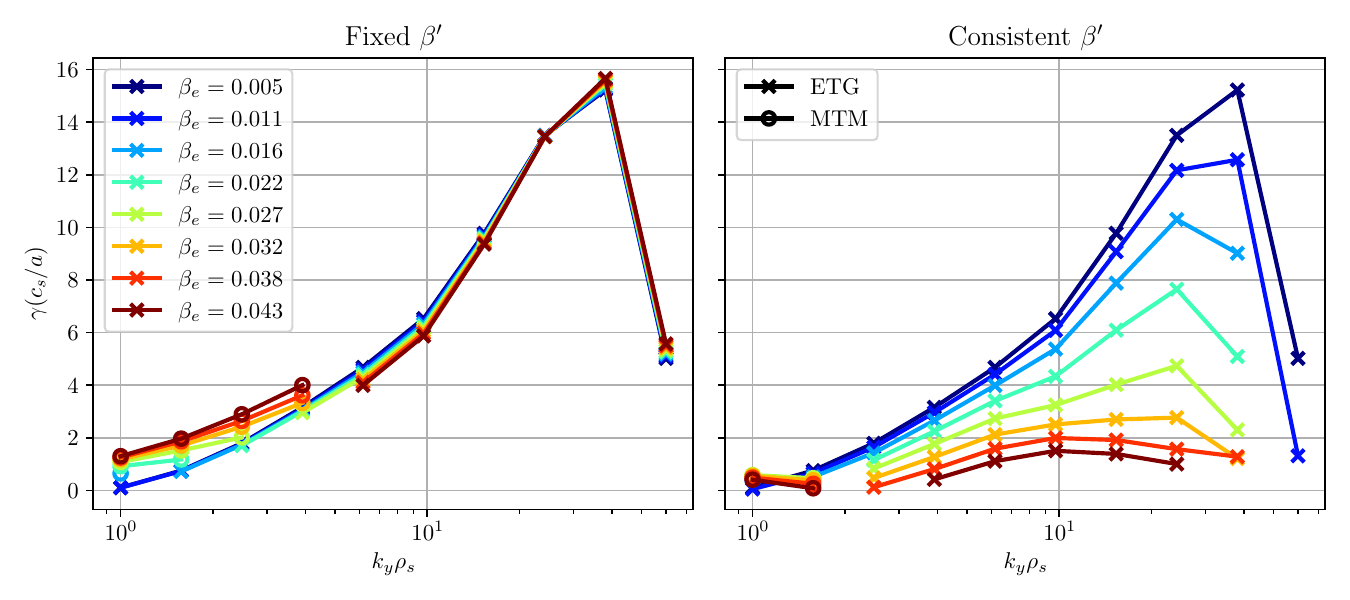}
    \caption{Sensitivity to $\beta_e$, with $\beta'$ held fixed (left) and varied self-consistently (right). ETG modes are shown by circles and MTM modes by crosses. The experimental value is $\beta_e = 0.005$.}
    \label{fig:linear_beta}
\end{figure}

ETG stabilisation with increasing $|\beta'|$ has been reported previously \cite{patel2022linear}. Figure \ref{fig:linear_beta} shows scans in $\beta_e$ for two cases: one with $\beta'$ held fixed, and another where $\beta'$ is varied self-consistently such that $\beta' = -\beta_e (p/p_e) (a/L_p)$. The lowest value, $\beta_e = 0.005$, corresponds to the experimental equilibrium. When $\beta'$ is fixed, ETG modes (circles) are largely unaffected, although a low-$k_y$ subdominant MTM branch (crosses) is destabilised at higher $\beta_e$\footnote{These modes can be distinguished using the linear tearing parameter $C_{\mathrm{tear}}$ \cite{patel2025impact}.}. In contrast, when $\beta'$ is varied consistently, the ETG modes are clearly stabilised. Including $B_{||}$ fluctuations produces similar qualitative behaviour, with a slight increase in instability at higher $\beta$, indicating that its inclusion is more important at larger $\beta$. Although fixed $\beta'$ scans suggest a possible role for MTMs, the required increase in $\beta$ is unrealistically large. When $\beta'$ is treated consistently, a clear scale separation between ETG and MTM modes appears.

\begin{figure}[!htb]
    \begin{subfigure}{0.49\textwidth}
        \centering
        \includegraphics[width=75mm]{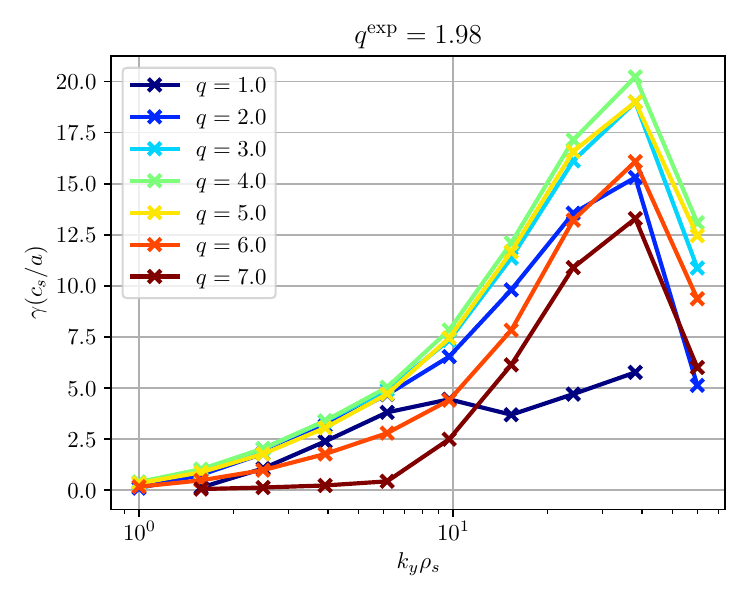}
        \caption{}
        \label{fig:48640_q_scan_ky}
    \end{subfigure}
    \begin{subfigure}{0.49\textwidth}
        \centering
        \includegraphics[width=75mm]{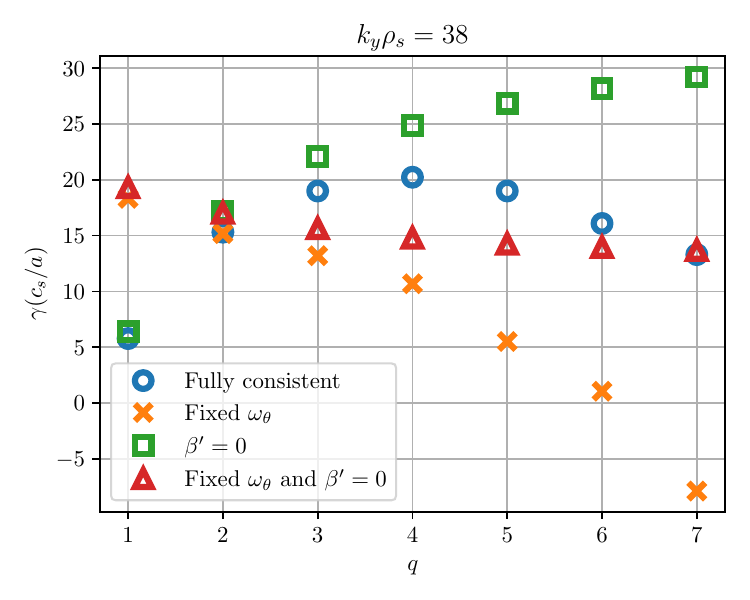}
        \caption{}
        \label{fig:48640_q_scan_stability}
    \end{subfigure}
    \begin{subfigure}{\textwidth}
        \centering
        \includegraphics[width=75mm]{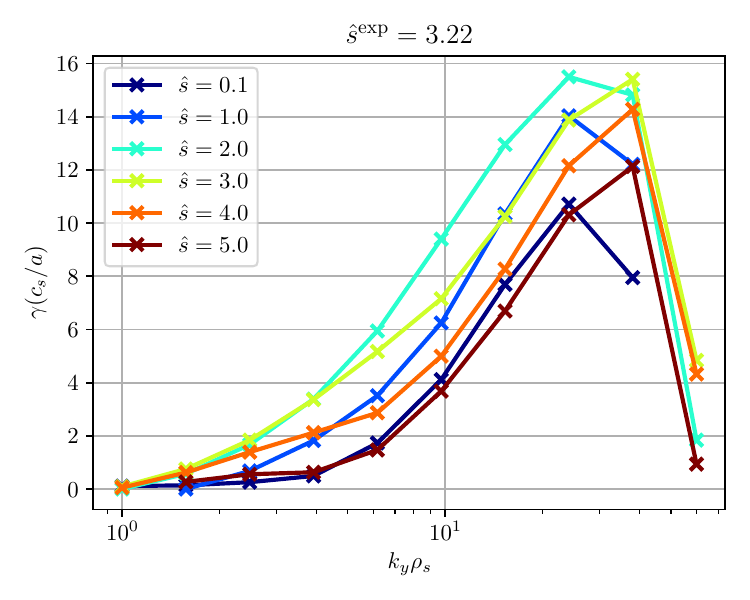}
        \caption{}
        \label{fig:48640_shat_scan_stability}
    \end{subfigure}
    \caption{(a) Linear sensitivity of ETG modes to $q$ over a range of $k_y\rho_s$. (b) Scans in $q$ at $k_y\rho_s = 38$, illustrating the roles of the parallel streaming term $\omega_\theta$ and the pressure component of $\omega_d$. (c) Linear sensitivity of ETG modes to magnetic shear $\hat{s}$.}
    \label{fig:q_scan}
\end{figure}

The dependence on safety factor $q$ is shown in Figure \ref{fig:48640_q_scan_ky}, where a clear non-monotonic behaviour is observed. As $q$ increases up to $q \approx 4$, the toroidal ETG modes become increasingly unstable, consistent with the critical gradient scaling $R/L_{T_e,\mathrm{crit}} \propto \hat{s}/q$ \cite{jenko2001critical}, with the higher $k_y$ modes being much more sensitive to $q$. However, for $q > 4$, the modes are stabilised. This behaviour can be understood by considering competing physical effects. The parallel streaming term stabilises toroidal ETG modes and scales as $\omega_\theta \propto 1/q$ (see equation 116 in \cite{candy2009unified}), suggesting that increasing $q$ should enhance instability. This trend is confirmed when $\omega_\theta$ is artificially held fixed (Figure \ref{fig:48640_q_scan_stability}); with the destabilising reduction in parallel streaming removed, increasing $q$ instead leads to stabilisation. The stabilisation at high $q$ is instead attributed to the pressure-gradient contribution to the magnetic drift frequency $\omega_d$ \cite{candy2016high}, which acts similarly to $\beta'$ stabilisation. When this contribution is removed by setting $\beta' = 0$, increasing $q$ leads only to destabilisation. This is equivalent to keeping $\alpha_{\mathrm{MHD}} \sim -q^2 \beta'$ constant. Finally, when $\omega_\theta$ is fixed and $\beta' = 0$, as shown in Figure \ref{fig:48640_q_scan_stability}, the growth rate becomes insensitive to $q$, indicating that these two mechanisms dominate the $q$ dependence. These results indicate that the critical gradient formula for ETG obtained in \cite{jenko2001critical} in the limit of large aspect ratio, low $\beta$ and low shaping is not applicable to finite $\beta$ ST plasmas.

The sensitivity to magnetic shear, $\hat{s}$, is shown in Figure \ref{fig:48640_shat_scan_stability}. A non-monotonic dependence is observed, with the peak growth rate occurring near $\hat{s} \approx 2$, in contrast to the simple scaling discussed above, due to the complicated relation between $\hat{s}$ and the drifts.

\subsubsection{Linear stability after the core rotation collapse}
%Plot of ETG turb on different surfaces after

\begin{figure}[!htb]
    \centering
    \includegraphics[width=150mm]{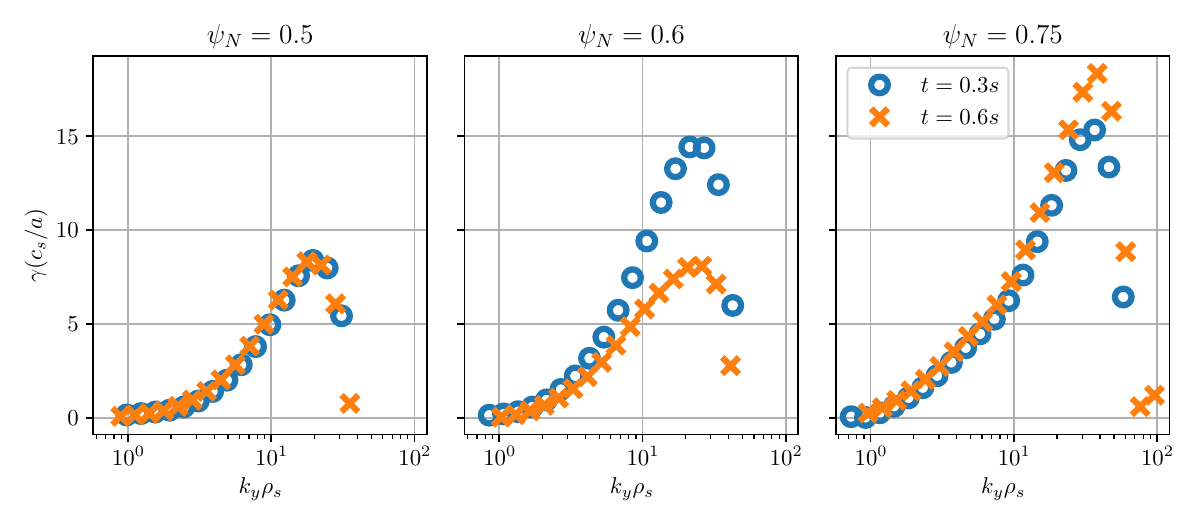}
    \caption{Comparison of linear growth rates before the core rotation collapse at $t = 0.3\,\mathrm{s}$ (blue circles) and after the collapse at $t = 0.6\,\mathrm{s}$ (orange crosses) for three flux surfaces: $\psi_N = 0.5$ (left), $\psi_N = 0.6$ (middle), and $\psi_N = 0.75$ (right). Note that $\psi_N = 0.6$ lies close to the $q = 3/2$ surface, where dominant MHD activity is expected, and gyrokinetic simulations may not be valid.}
    \label{fig:48640_etg_time_comp}
\end{figure}

Figure \ref{fig:48640_etg_time_comp} compares the linear stability properties at three flux surfaces, both before the core rotation collapse ($t = 0.3\,\mathrm{s}$) and after the collapse ($t = 0.6\,\mathrm{s}$). At $\psi_N = 0.5$, similar growth rates are observed at both times, indicating that the equilibrium changes accompanying the collapse do not significantly alter the linear stability on inner flux surfaces, consistent with the near-unchanged $a/L_{T_e}$. At $\psi_N \approx 0.6$, where $\gamma_\mathrm{E \times B}$ is close to zero, a reduction in $\chi_e^{\mathrm{anom}}$ is observed (Figure \ref{fig:48640_chi_ratio}). However, this location lies close to the $q = 3/2$ surface, where large-scale MHD activity is expected, making it difficult to disentangle the relative contributions of large-scale MHD and small-scale turbulence. Nevertheless, a reduction in the linear growth rate is also observed at this surface. At $\psi_N = 0.75$, a larger $a/L_{T_e}$ leads to increased growth rates. Nonlinear simulations will show that ETG transport at this radius is suppressed by the higher $\gamma_\mathrm{E \times B}$, consistent with the steeper gradient. In contrast, towards the core, similar growth rates are observed despite a significant reduction in $\gamma_\mathrm{E \times B}$, and it will be shown that ETG modes contribute negligibly to the anomalous transport in this region. Given the strong linear sensitivity to $a/L_{T_e}$, we now investigate whether these features persist in nonlinear simulations and whether the predicted fluxes are consistent with experimental observations.

\subsection{Nonlinear sensitivities in L-mode}
This section compares the nonlinear heat flux predictions to the experiment for both an outer ($\psi_N=0.75$) and an inner ($\psi_N=0.5$) surface before ($t=0.3\,\mathrm{s}$) and after ($t=0.6\,\mathrm{s}$) the core rotation collapse. Heat fluxes are stated in gyroBohm units defined as $Q_{gB} = n_e T_e c_s (\rho_s/a)^2$. The stiffness of the transport is examined through scans in $a/L_{T_e}$ along with the sensitivity to $\gamma_\mathrm{E \times B}$. Nonlinear simulations with GPU-CGYRO have been performed, utilising 32 points along the field line and 16 pitch angles for a reduced computational cost, which was found to have little impact on the linear physics. The wavenumber spectrum utilised is stated at the start of each case.

\subsubsection{Outer surface \texorpdfstring{$\psi_N=0.75$}{psiN = 0.75} before the rotation collapse}

Nonlinear simulations were performed with a $k_y$ spectrum chosen to fully capture the ETG range. This required 96 binormal wavenumbers, with a minimum value of $k_{y,\min}\rho_s = 0.8$. Convergence tests indicated that an extended radial domain was necessary to resolve the radial streamers that develop in these simulations. Consequently, $k_{x,\min}\rho_s \sim 0.2$ was used with 768 radial wavenumbers.

\begin{figure}[!htb]
    \centering
     \includegraphics[width=150mm]{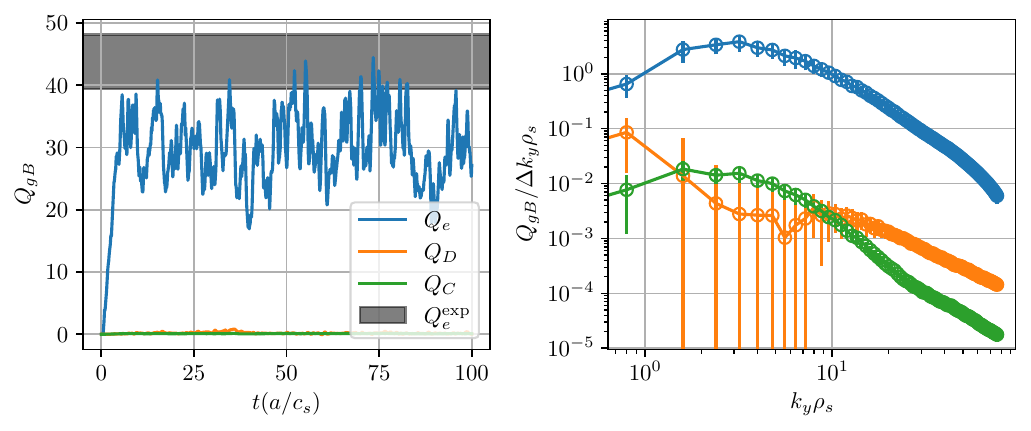}
    \caption{Nonlinear simulation of the outer surface at $\psi_N=0.75$ at $t=0.3\,\mathrm{s}$ in \#48640. Left: time trace of the heat flux, separated by species. The grey bar indicates the experimental level of anomalous heat flux obtained from TRANSP. Right: $k_y$ heat flux spectrum averaged over the latter half of the simulation.}
    \label{fig:48640_t300_psi075_nl_flux}
\end{figure}

Figure \ref{fig:48640_t300_psi075_nl_flux} shows the time evolution of the heat fluxes, along with the $k_y$ spectrum averaged over the latter half of the simulation. The heat transport is dominated by the electron channel, while the ion channels (deuterium and carbon) remain close to zero. The peak heat flux occurs at $k_y\rho_s = 2.4$, which is within a factor of two of the peak of the $\gamma / k_y^2$ spectrum (often used in mixing-length estimates) at $k_y\rho_s = 1.2$. This suggests that $\gamma / k_y^2$ is a useful linear proxy for the nonlinear flux spectrum, although the nonlinear flux peak is shifted to higher $k_y$, as commonly observed in reduced ETG transport models \cite{hatch2022reduced}.

\begin{figure}[!htb]
    \begin{subfigure}{0.99\textwidth}
        \centering
        \includegraphics[width=150mm]{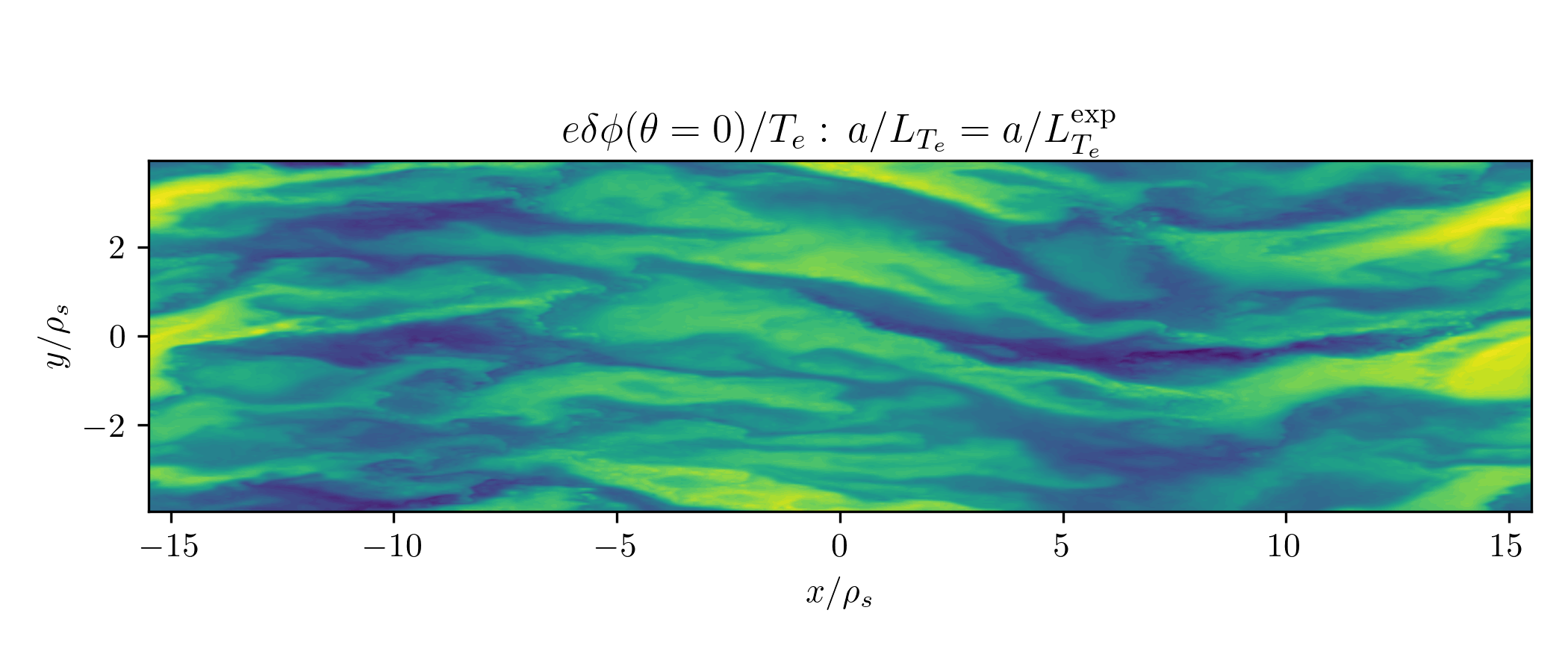}
        \caption{}
        \label{fig:48640_t300_psi075_base_nl_phi}
    \end{subfigure}
    \begin{subfigure}{0.99\textwidth}
        \centering
        \includegraphics[width=150mm]{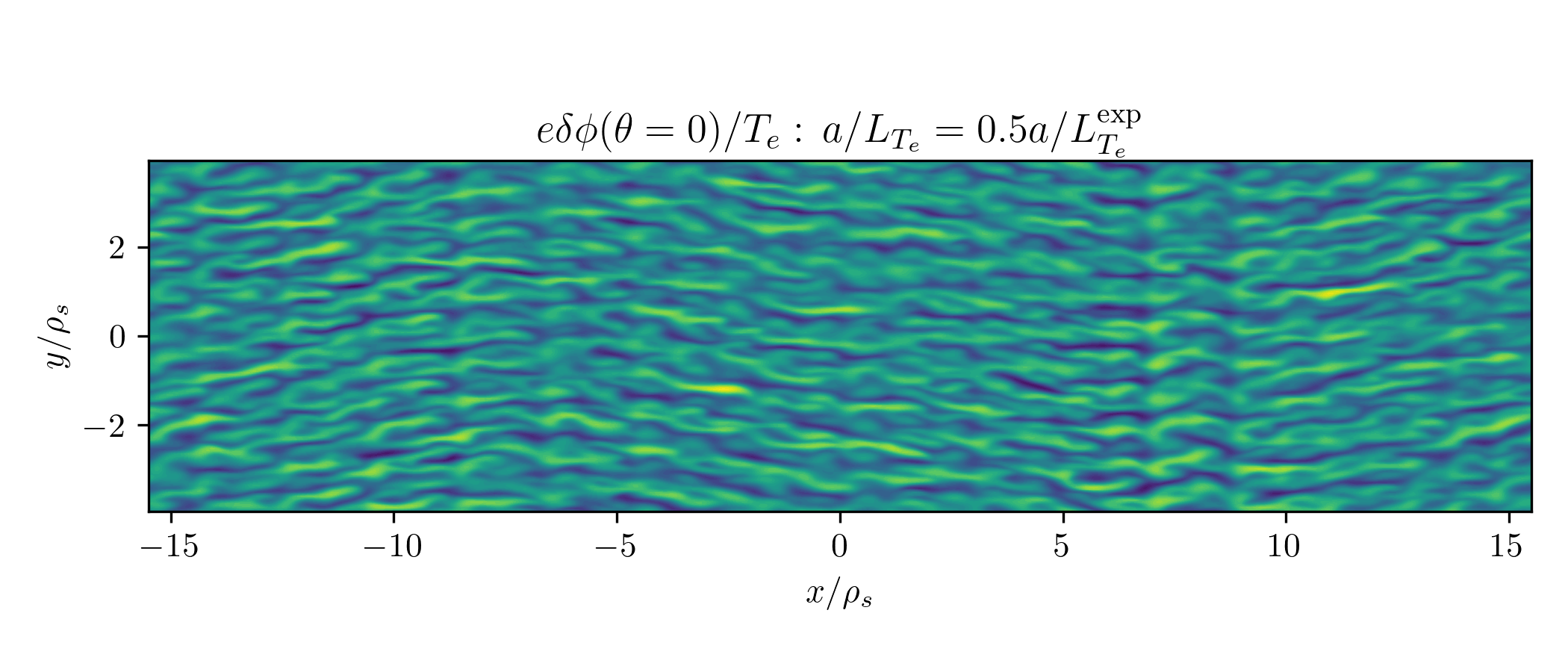}
        \caption{}
        \label{fig:48640_t300_psi075_alte_050_nl_phi}
    \end{subfigure}
    \caption{Real-space electrostatic potential fluctuations $\phi$ at the final simulation time slice at the outboard mid-plane ($\theta=0$). (a) Simulation using the experimental $a/L_{T_e}$. (b) Simulation with $a/L_{T_e}$ reduced by 50\%. At the experimental $a/L_{T_e}$, large radial structures form, generating significant transport, but as the drive to the ETG is reduced, the radial extent of the fluctuations reduces, lowering the level of transport.}
    \label{fig:48640_t300_psi075_nl_phi}
\end{figure}

Figure \ref{fig:48640_t300_psi075_base_nl_phi} shows the electrostatic potential fluctuations in real space at $\theta = 0$. Radially extended streamers are clearly visible and are responsible for the large transport levels. If the radial domain is too small, these streamers can effectively ``short-circuit'' the simulation, artificially increasing the flux by orders of magnitude.

\begin{figure}[!htb]
    \begin{subfigure}{0.49\textwidth}
        \centering
        \includegraphics[width=75mm]{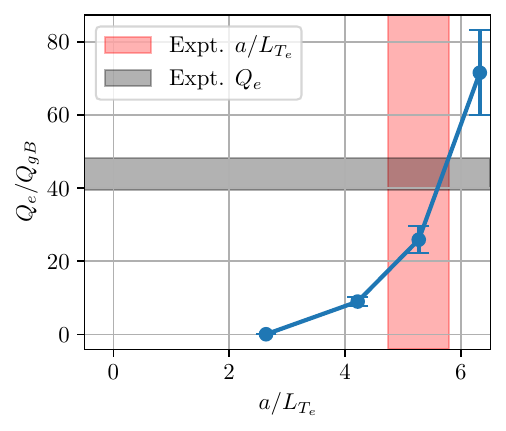}
        \caption{}
        \label{fig:48640_t300_psi075_alte}
    \end{subfigure}
    \begin{subfigure}{0.49\textwidth}
        \centering
        \includegraphics[width=75mm]{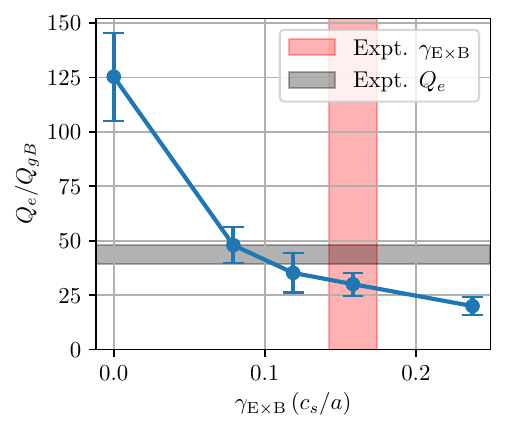}
        \caption{}
        \label{fig:48640_t300_psi075_game}
    \end{subfigure}
    \caption{Nonlinear simulations at $\psi_N=0.75$ and $t=0.3\,\mathrm{s}$ in \#48640 showing the sensitivity of electron heat flux to (a) $a/L_{T_e}$ and (b) $\gamma_\mathrm{E \times B}$. The horizontal grey bar indicates the experimental estimate of $Q_e$, while the vertical red bar denotes the uncertainty in the scanned parameter obtained from uncertainty in the fit to the Thomson scattering data.}
    \label{fig:48640_t300_psi075_flux}
\end{figure}

The experimental electron heat flux obtained from TRANSP is indicated by the grey bar in Figure \ref{fig:48640_t300_psi075_nl_flux}. The simulation under-predicts the electron transport by approximately $40\%$. Given the uncertainties inherent in comparing experimental measurements with interpretive modelling, it is important to assess the sensitivity of the results to input parameters. In stiff transport regimes, small changes in growth rates can lead to large variations in flux. This sensitivity is illustrated by a scan in $a/L_{T_e}$ (Figure \ref{fig:48640_t300_psi075_alte}), where a $20\%$ increase in the electron temperature gradient leads to a doubling of the predicted heat flux. The heat flux exhibits approximately cubic scaling, as observed in JET ETG pedestal experiments \cite{chapman2022role} and in both slab and toroidal ETG simulations \cite{adkins2022electromagnetic, adkins2023scale, adkins2026asymptotic}. The red bar denotes the experimental uncertainty, which is calculated from the uncertainty in the fit to the Thomson scattering data. Within experimental uncertainties, this brings the simulation into agreement with observations. Accurately matching experimental measurements in gradient-driven simulations is challenging. However, flux-driven approaches often yield better agreement, particularly in regimes with strong sensitivity. Reducing the drive leads to a substantial decrease in flux, accompanied by a reduction in the radial extent of the streamers (Figure \ref{fig:48640_t300_psi075_alte_050_nl_phi}).

The sensitivity to $\gamma_\mathrm{E \times B}$ is shown in Figure \ref{fig:48640_t300_psi075_game}. Increasing $\gamma_\mathrm{E \times B}$ produces a stabilising effect, although the impact is less pronounced than for $a/L_{T_e}$. When $\gamma_\mathrm{E \times B}$ is reduced to zero, a large increase in heat flux is observed, associated with more radially extended streamers. Matching the experimental flux would require a $\sim 50\%$ reduction in $\gamma_\mathrm{E \times B}$, which lies well outside the experimental uncertainty range.

\subsubsection{Outer surface \texorpdfstring{$\psi_N=0.75$}{psiN = 0.75} after the rotation collapse}

\begin{figure}[!htb]
    \begin{subfigure}{0.49\textwidth}
        \centering
        \includegraphics[width=75mm]{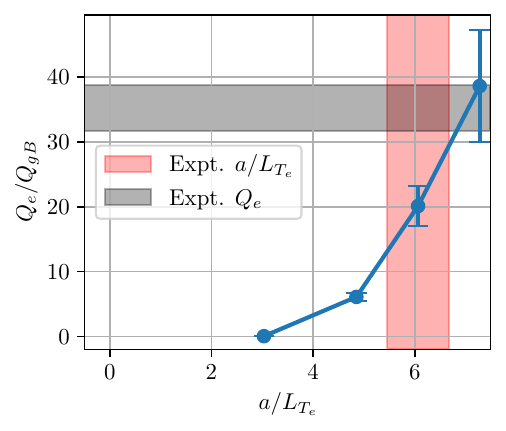}
        \caption{}
        \label{fig:48640_t600_psi075_alte}
    \end{subfigure}
    \begin{subfigure}{0.49\textwidth}
        \centering
        \includegraphics[width=75mm]{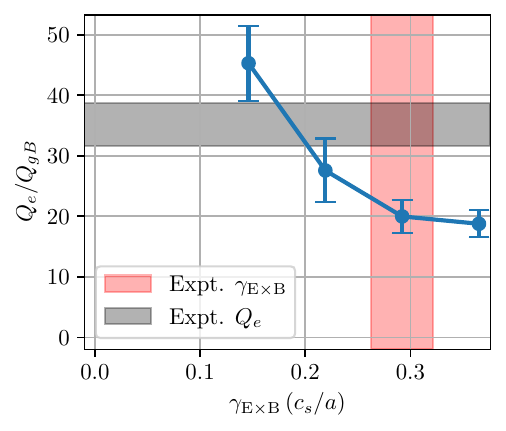}
        \caption{}
        \label{fig:48640_t600_psi075_game}
    \end{subfigure}
    \caption{Nonlinear simulations at $\psi_N = 0.75$ and $t = 0.6\,\mathrm{s}$ in \#48640, showing the sensitivity of electron heat flux to (a) $a/L_{T_e}$ and (b) $\gamma_\mathrm{E \times B}$. The horizontal grey bar indicates the experimental estimate of $Q_e$, while the vertical red bar denotes the uncertainty in the scanned parameter.}
    \label{fig:48640_t600_psi075_flux}
\end{figure}

Equivalent nonlinear simulations were performed after the core rotation collapse at $t = 0.6\,\mathrm{s}$, as shown in Figure \ref{fig:48640_t600_psi075_flux}. At this time, both $\gamma_\mathrm{E \times B}$ and $a/L_{T_e}$ are higher than in the $t = 0.3\,\mathrm{s}$ case. The baseline simulation again under-predicts the electron heat flux relative to the experiment, but increasing $a/L_{T_e}$ destabilises the ETG modes. The sensitivity is slightly reduced compared to the earlier time, which is consistent with the stabilising influence of the larger $\gamma_\mathrm{E \times B}$. A similarly strong dependence on the $\mathrm{E \times B}$ shear is observed, as at $t = 0.3\,\mathrm{s}$. Within the uncertainties of the measurements, the experimental heat flux can be reproduced, indicating that ETG turbulence can account for the observed anomalous transport at this surface. Nonetheless, additional sources of uncertainty, such as the assumed $Z_\mathrm{eff}$, remain important, as linear studies have shown its significant impact on ETG stability.

\subsubsection{Inner surface \texorpdfstring{$\psi_N=0.5$}{psiN = 0.5} before the rotation collapse}

A similar approach was applied to the inner flux surface, where the ETG modes are more stable and exist over a narrower range of $k_y$, reducing the resolution requirements for the simulations. Specifically, 48 binormal wavenumbers with $k_{y,\min}\rho_s = 0.8$ were sufficient to capture the linear range of unstable modes, while 384 radial wavenumbers with $k_{x,\min}\rho_s \sim 0.2$ adequately resolved the radial structures.

\begin{figure}[!htb]
    \centering
    \includegraphics[width=150mm]{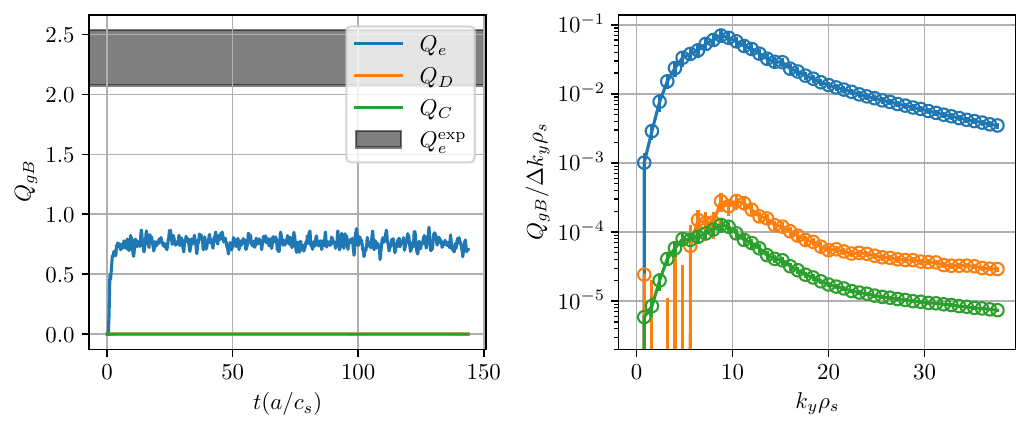}
    \caption{Nonlinear simulation of the inner surface at $\psi_N = 0.5$ at $t = 0.3\,\mathrm{s}$ in \#48640. Left: time trace of the heat flux, separated by species. The grey bar indicates the experimental level of anomalous heat flux. Right: $k_y$ flux spectrum averaged over the latter half of the simulation.}
    \label{fig:48640_t300_psi_05_nl_flux}
\end{figure}

\begin{figure}[htb]
    \begin{subfigure}{0.49\textwidth}
        \centering
        \includegraphics[width=75mm]{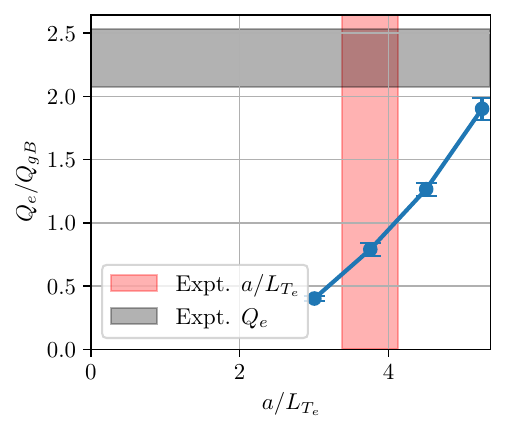}
        \caption{}
        \label{fig:48640_t300_psi05_alte}
    \end{subfigure}
    \begin{subfigure}{0.49\textwidth}
        \centering
        \includegraphics[width=75mm]{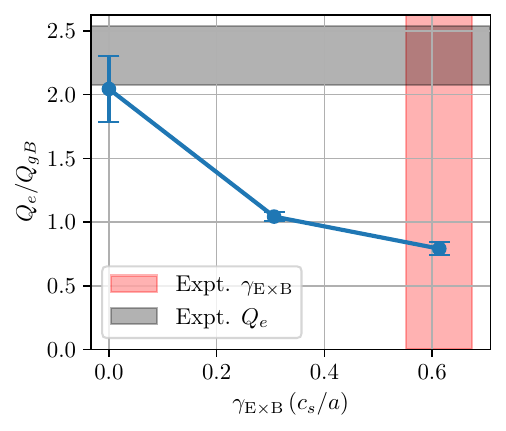}
        \caption{}
        \label{fig:48640_t300_psi05_game}
    \end{subfigure}
    \caption{Nonlinear simulations at $\psi_N = 0.5$ and $t = 0.3\,\mathrm{s}$ in \#48640, showing the sensitivity of electron heat flux to (a) $a/L_{T_e}$ and (b) $\gamma_\mathrm{E \times B}$. The horizontal grey bar indicates the experimental estimate of $Q_e$, while the vertical red bar shows the uncertainty in the scanned parameter.}
    \label{fig:48640_t300_psi05_flux}
\end{figure}

Figure \ref{fig:48640_t300_psi_05_nl_flux} shows the nonlinear electron heat flux using the baseline parameters. The predicted electron flux is significantly lower than the experimental estimate, with a peak in the $k_y$ spectrum at $k_y\rho_s = 8.8$, substantially higher than the peak observed at $\psi_N = 0.75$. Sensitivity scans similar to those performed for the outer surface indicate that the electron heat flux is much less responsive to changes in $a/L_{T_e}$ (Figure \ref{fig:48640_t300_psi05_alte}). A 50\% increase in $a/L_{T_e}$ is required to achieve agreement with the experimental flux, well beyond the experimental uncertainty of $\sim 10\%$. Similarly, reducing the $\mathrm{E \times B}$ shear has only a weak effect (Figure \ref{fig:48640_t300_psi05_game}); the ETG mode transport matches the experimental heat flux only when $\gamma_\mathrm{E \times B}$ is nearly zero.

\begin{figure}[!htb]
    \begin{subfigure}{\textwidth}
        \centering
        \includegraphics[width=150mm]{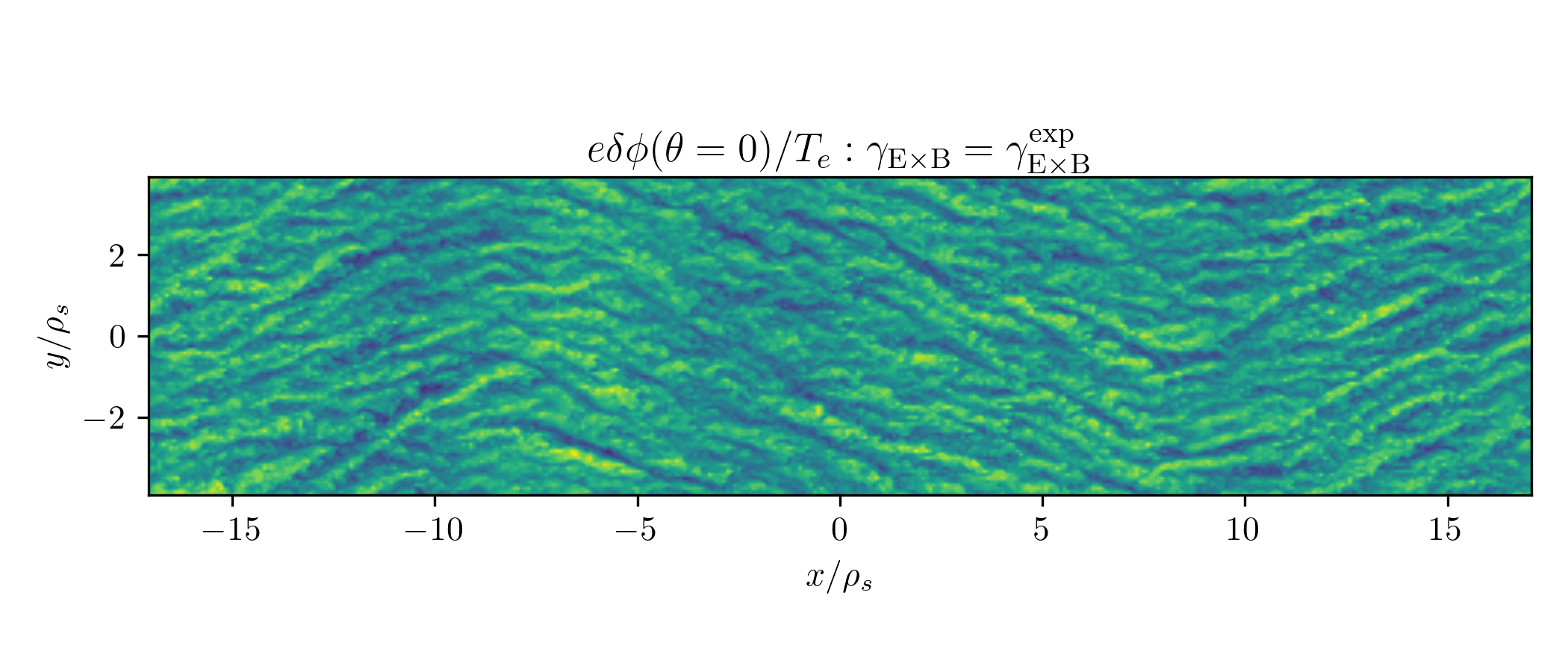}
        \caption{}
        \label{fig:48640_t300_psi05_phi}
    \end{subfigure}
    \begin{subfigure}{\textwidth}
        \centering
        \includegraphics[width=150mm]{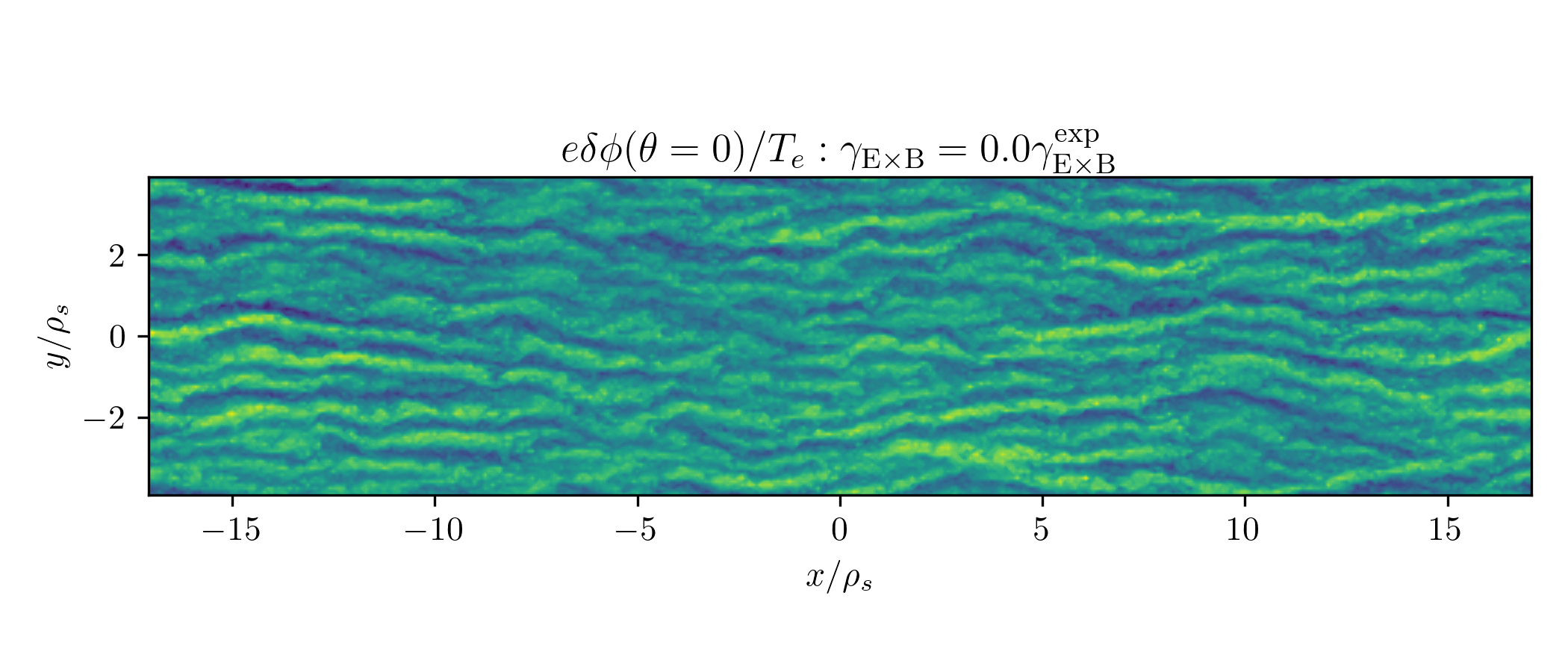}
        \caption{}
        \label{fig:48640_t300_psi05_gam_000phi}
    \end{subfigure}
    \caption{Real-space electrostatic potential fluctuations $\phi$ at the outboard mid-plane ($\theta = 0$) at the final simulation time slice for the inner surface at $\psi_N=0.5$ at $t=0.3\,\mathrm{s}$ in \#48640. (a) Simulation with the experimental $\gamma_\mathrm{E \times B}$. (b) Simulation with $\gamma_\mathrm{E \times B} = 0$.}
\end{figure}

Figure \ref{fig:48640_t300_psi05_phi} shows the electrostatic potential fluctuations at this surface. The radial extent of the streamers is significantly reduced compared to the outer surface, and a finer structure is present in the binormal direction, consistent with the peak heat flux occurring at higher $k_y$. A weak zonal flow structure, driven by $\gamma_\mathrm{E \times B}$, shears the streamers apart\footnote{Note that CGYRO implements $\mathrm{E \times B}$ shear via a box scale zonal mode}. Figure \ref{fig:48640_t300_psi05_gam_000phi} shows the electrostatic potential when $\gamma_\mathrm{E \times B}$ is removed, with the streamers elongating radially, lowering the radial wavenumber and increasing the heat flux by roughly a factor of three.

These results suggest that ETG modes alone are insufficient to account for the level of anomalous electron heat transport predicted by TRANSP on this inner surface. This motivates further investigation of ion-scale turbulence, which becomes increasingly relevant as the radial location moves inward, consistent with the relative increase in ion heat transport observed in Figure \ref{fig:48640_chi_ratio}.

\subsubsection{Inner surface \texorpdfstring{$\psi_N=0.5$}{psiN = 0.5} after the rotation collapse}

A similar analysis was conducted for the inner surface after the core rotation collapse at $t = 0.6\,\mathrm{s}$. The same $k_y$ range was used as in simulations prior to rotation collapse, but 384 radial modes with $k_{x,\min}\rho_s \sim 0.28$ were required due to the reduced radial extent of the ETG fluctuations. At this time, $\gamma_\mathrm{E \times B}$ is significantly reduced from $0.61$ to $-0.05\,c_s/a$, but $a/L_{T_e}$ is largely unaffected going from $a/L_{T_e}=3.76$ to $3.78$. 

\begin{figure}[!htb]
    \begin{subfigure}{0.32\textwidth}
        \centering
        \includegraphics[width=50mm]{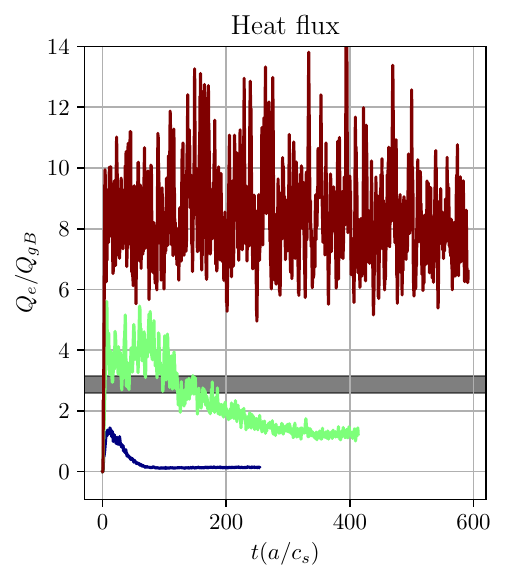}
        \caption{}
        \label{fig:48640_t600_psi05_alte_zonal_flux}
    \end{subfigure}
    \begin{subfigure}{0.32\textwidth}
        \centering
        \includegraphics[width=50mm]{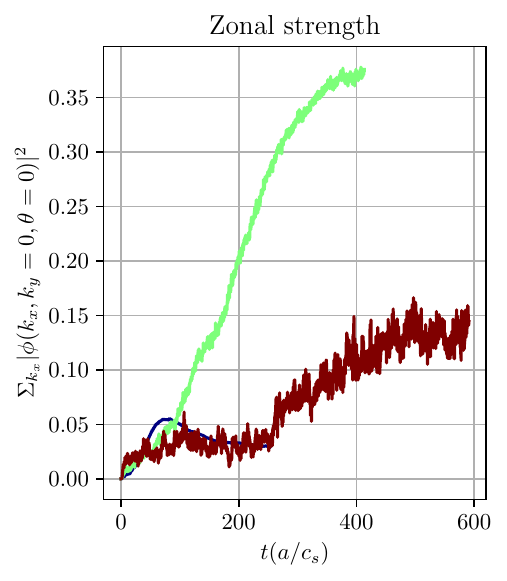}
        \caption{}
        \label{fig:48640_t600_psi05_alte_zonal_strength}
    \end{subfigure}
    \begin{subfigure}{0.32\textwidth}
        \centering
        \includegraphics[width=50mm]{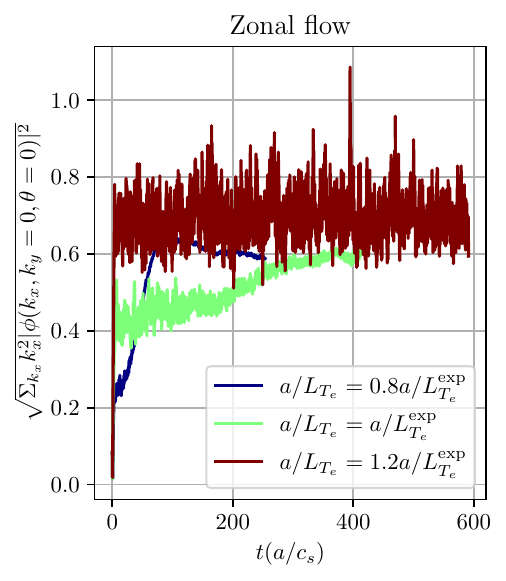}
        \caption{}
        \label{fig:48640_t600_psi05_alte_zonal_flow}
    \end{subfigure}
    \caption{Nonlinear simulations of the inner surface at $\psi_N=0.5$ at $t=0.6\,\mathrm{s}$ in \#48640. A scan in $a/L_{T_e}$ showing (a) the electron heat flux, (b) zonal strength, and (c) root-mean square zonal flow as a function of time. Note that $\gamma_\mathrm{E \times B}^\mathrm{exp} = -0.05\,c_s/a$ was utilised for these simulations.}
    \label{fig:48640_t600_psi05_alte_zonal}
\end{figure}

Figure \ref{fig:48640_t600_psi05_alte_zonal_flux} shows the time evolution of the electron heat flux for three simulations with varying $a/L_{T_e}$. For the experimental gradient, the simulation initially saturates at a level comparable to the experimental flux, similar to the $t = 0.3\,\mathrm{s}$ simulation without $\mathrm{E \times B}$ shear. Over a longer timescale the heat flux decreases, saturating below the experimental value with a strong zonal mode forming\footnote{Simulations where this strong zonal mode formed required careful dealiasing in the parallel direction to avoid grid-scale oscillations, which does not occur in other simulations; see implementation outlined in \cite{candy2026skew}}. This slow zonal build-up has been observed previously by G. Colyer \textit{et al.} \cite{colyer2017collisionality}, who identified long-timescale zonal fluctuations eventually dominating the simulation.

\begin{figure}[!htb]
    \begin{subfigure}{\textwidth}
        \centering
        \includegraphics[width=150mm]{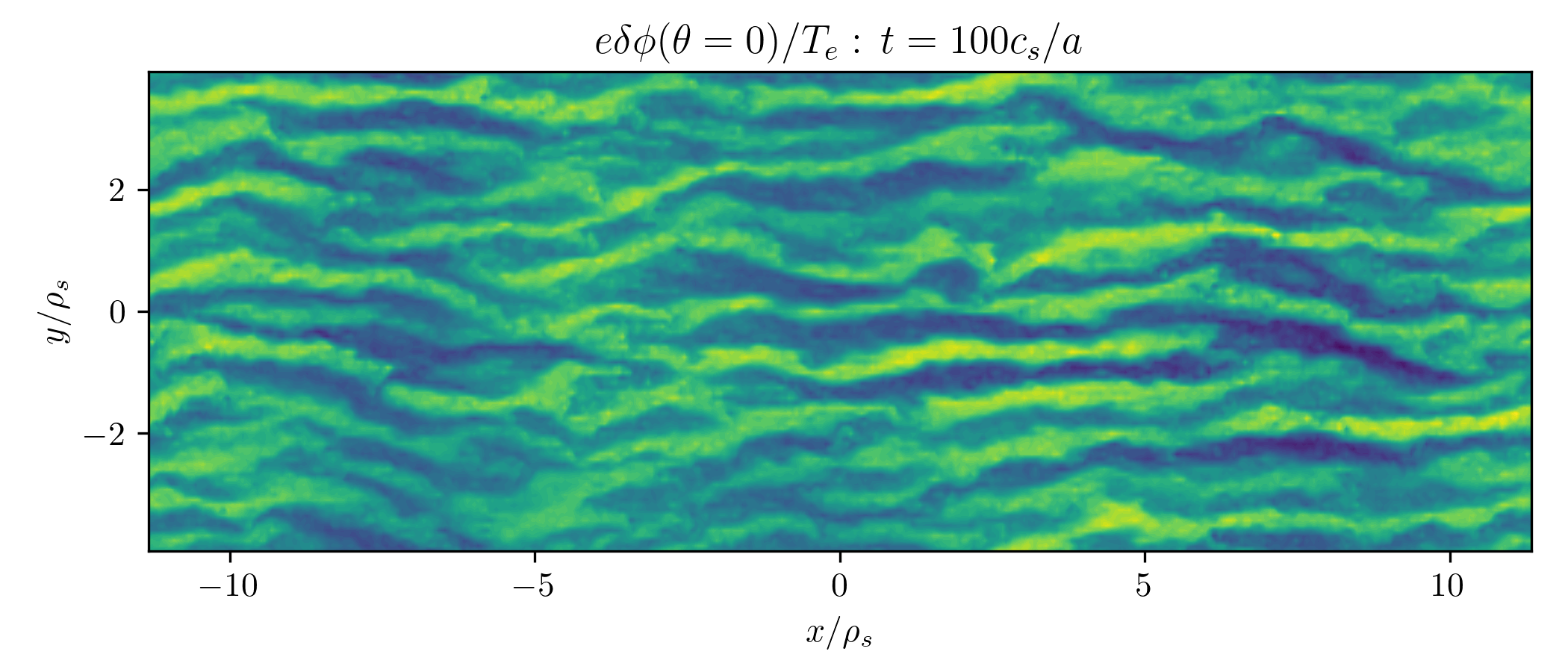}
        \caption{}
        \label{fig:48640_t600_psi05_phi_initial}
    \end{subfigure}
    \begin{subfigure}{\textwidth}
        \centering
        \includegraphics[width=150mm]{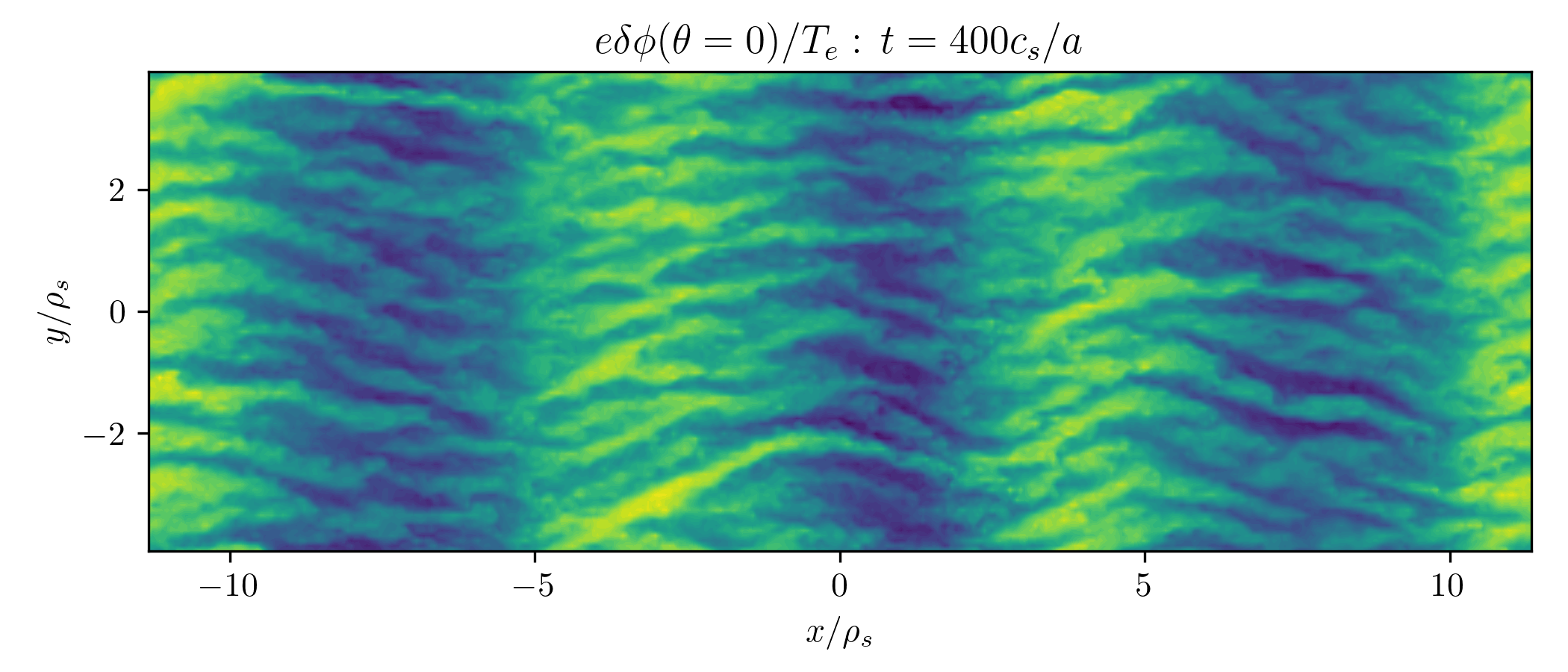}
        \caption{}
        \label{fig:48640_t600_psi05_phi_final}
    \end{subfigure}
    \caption{Real-space electrostatic potential fluctuations $\phi$ at the outboard mid-plane ($\theta=0$) for the inner surface at $\psi_N=0.5$ after the rotation collapse. (a) $t = 100\,a/c_s$ (quasi-saturated phase), (b) $t = 400\,a/c_s$ (final saturated state)  in \#48640.}
    \label{fig:48640_t600_psi05_phi}
\end{figure}

Figures \ref{fig:48640_t600_psi05_alte_zonal_strength} and \ref{fig:48640_t600_psi05_alte_zonal_flow} show the evolution of the zonal strength, defined as $\sum_{k_x}|\phi(k_x, k_y=0, \theta=0)|^2$, and the root-mean square zonal flow, $\sqrt{\sum_{k_x} k_x^2 |\phi(k_x, k_y=0, \theta=0)|^2}$, respectively. In the baseline simulation, the zonal strength grows over time, beginning to saturate around $t = 400\,a/c_s$ (corresponding to an electron thermal time $\sim 3.4\times 10^4\,a/v_{th,e}$), while the heat flux reaches its new quasi-saturated state slightly earlier at $t \sim 300\,a/c_s$. Similar to the heat flux, the zonal flow quasi-saturates at early times around $0.45\,c_s$ but increases during the heat flux collapse and saturates at $0.6\,c_s$, shown in Figure \ref{fig:48640_t600_psi05_alte_zonal_flow}. These two phases correspond to turbulence-driven and self-driven zonal flows respectively. Turbulence-driven zonal flows are generated through nonlinear interactions within the drift-wave turbulence and are therefore present during the quasi-saturated turbulent state. In contrast, self-driven zonal flows refer to the zonal mode growing through its own nonlinear drive, allowing the zonal flow to continue growing after the turbulence has been suppressed. The real-space electrostatic potential fluctuations clearly illustrate this behaviour. Figures \ref{fig:48640_t600_psi05_phi_initial} and \ref{fig:48640_t600_psi05_phi_final} show the fluctuations during the initial quasi-saturated phase at $t = 100\,a/c_s$ and the final saturated state at $t = 400\,a/c_s$, respectively. Initially, the radial streamers resemble those seen before the rotation collapse, but once full saturation is reached, distinct zonal bands form, suppressing radial transport.

Further examination of Figure \ref{fig:48640_t600_psi05_alte_zonal_flux} shows that simulations with $a/L_{T_e}$ reduced by 20\% exhibit similar qualitative behaviour. However, the reduction in heat flux occurs more rapidly, beginning at $t=20\,a/c_s$, and the zonal strength reaches a lower level during the quasi-saturated state. The zonal flow initially quasi-saturates at $0.28\,c_s$ (turbulence-driven), before eventually fully saturating at $0.6\,c_s$ (self-driven), similar to the amplitude in the baseline scenario, suggesting that changing the linear drive does not affect the saturated zonal flow amplitude. Conversely, a 20\% increase in $a/L_{T_e}$ shows a slow, steady growth of zonal strength, which saturates at a lower level by $t \sim 500\,a/c_s$. However, this initial turbulence-driven zonal flow saturates at $0.7\,c_s$, above the previously observed self-driven zonal flow. Following the analysis of G. Colyer \textit{et al.} \cite{colyer2017collisionality}, this can be understood as a competition between the linear drive and zonal-nonzonal interactions: lower linear drive requires less zonal amplitude to achieve suppression. In the quasi-saturated phases, all three simulations exhibit turbulence-driven zonal flows. Although the turbulence-generated zonal flows in the quasi-saturated phase increase with the gradient, the self-driven zonal flows remain constant at $\sim0.6\, c_s$. The turbulence-driven flows in the two lower-gradient simulations remain below the self-driven flow amplitude, and saturation is observed. However, in the higher-gradient case, the turbulence-driven flows ($\sim0.7c_s$) exceed the level seen in the self-driven flows, making suppression less certain. The existence of a drive threshold separating a zonally dominated, low-transport state from a strongly driven one is qualitatively analogous to the Dimits shift established for ion-scale ITG turbulence \cite{dimits2000comparisons, ivanov2020zonally, ivanov2022dimits}. We emphasise that those studies concern ITG turbulence at ion scales, where the zonal dynamics differ from the electron-scale case considered here in which the ion response is adiabatic; the correspondence should therefore be regarded as qualitative.

\begin{figure}[!htb]
    \begin{subfigure}{0.32\textwidth}
        \centering
        \includegraphics[width=50mm]{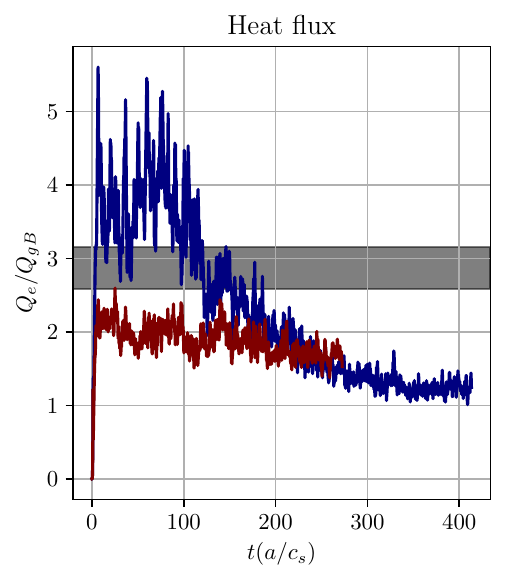}
        \caption{}
        \label{fig:48640_t600_psi05_shat_zonal_flux}
    \end{subfigure}
    \begin{subfigure}{0.32\textwidth}
        \centering
        \includegraphics[width=50mm]{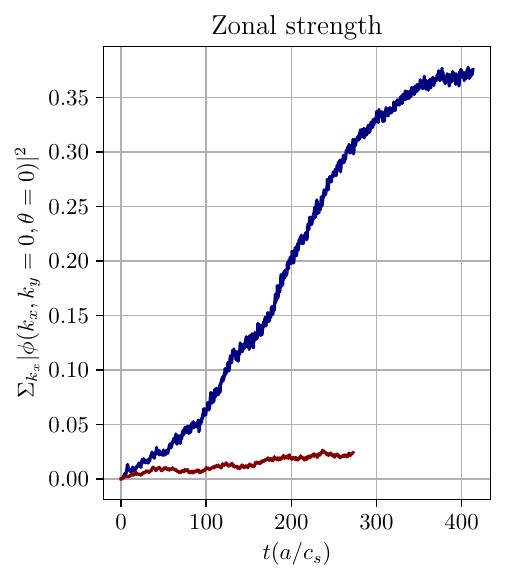}
        \caption{}
        \label{fig:48640_t600_psi05_shat_zonal_strength}
    \end{subfigure}
    \begin{subfigure}{0.32\textwidth}
        \centering
        \includegraphics[width=50mm]{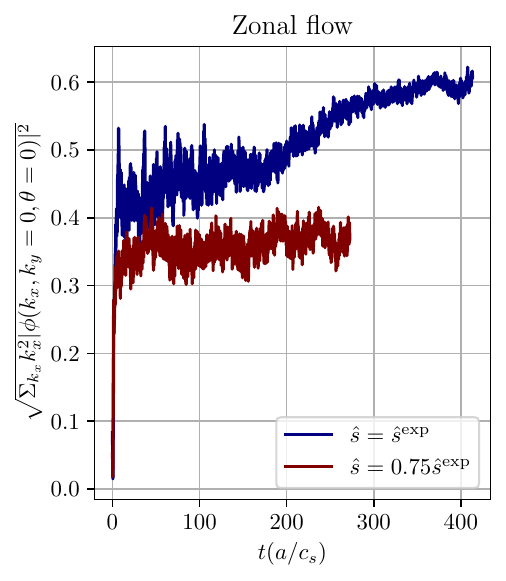}
        \caption{}
        \label{fig:48640_t600_psi05_shat_zonal_flow}
    \end{subfigure}
    \caption{Nonlinear simulations of the inner surface at $\psi_N=0.5$ at $t=0.6\,\mathrm{s}$ in \#48640. A scan in $\hat{s}$ showing (a) electron heat flux, (b) zonal strength, and (c) root-mean square zonal flow as a function of time. This demonstrates that the balance between the zonal flow stabilisation and gradient drive is sensitive to magnetic shear.}
    \label{fig:48640_t600_psi05_shat_zonal}
\end{figure}

The dependence of this phenomenon on magnetic shear was explored by reducing $\hat{s}$ by 25\% to better match the value at $t = 0.3\,\mathrm{s}$ while keeping the same radial box and resolution. Figure \ref{fig:48640_t600_psi05_shat_zonal} shows the heat flux, zonal strength, and root-mean square zonal flow for these cases. While linear stability remained similar, the zonal growth rate was substantially slower. The lower-$\hat{s}$ case would require much longer times to reach a zonal amplitude capable of suppressing transport, explaining why this behaviour is less evident at earlier times. The reduced zonal growth rate at lower $\hat{s}$ is consistent with a weaker nonlinear drive of the zonal modes. Higher magnetic shear increases the rate at which energy is transferred from the streamer-like nonzonal modes into the zonal channel, accelerating the transition to the zonal-dominated state. Since the linear drive is similar in both cases, the difference in zonal evolution is primarily nonlinear in origin, consistent with the framework of G. Colyer \textit{et al.}, in which the zonal build-up rate is set by the competition between nonlinear excitation and collisional damping.

% \begin{figure}[!htb]
%     \centering
%     \includegraphics[width=0.99\textwidth]{figures/48640_psin_05_t_600_shat_zonal_scan.pdf}
%     \caption{Nonlinear simulations of the inner surface at $\psi_N=0.5$ at $t=0.6\,\mathrm{s}$ in \#48640. A scan in $\hat{s}$ showing (a) electron heat flux, (b) zonal strength, and (c) squared zonal flow as a function of time. This shows that the balance between the zonal flow stabilisation and gradient drive is sensitive. }
%     \label{fig:48640_t600_psi05_shat_zonal}
% \end{figure}

Returning to the other cases, a slow zonal growth is occasionally observed in the outer-surface simulations and at $\psi_N=0.5$ before the collapse, but the growth rate is typically very small. Capturing this fully would require simulation times beyond practical computational limits. Additionally, if it were possible to run the simulation for such long timescales, ion-scale transport may become relevant, necessitating true multi-scale simulations to accurately model the system.

\subsection{L-mode summary}
Overall, ETG modes become increasingly relevant when moving from the inner core to the outer core. Within $\psi_N \lesssim 0.4$, ETG modes are generally stable. Around $\psi_N \sim 0.5$, ETG modes can drive non-negligible transport in the presence of significant $\mathrm{E \times B}$ shear, although the predicted flux remains below experimental levels. However, when the $\mathrm{E \times B}$ shear is reduced, for example following an MHD-induced rotation collapse, ETG-driven transport increases and becomes experimentally relevant. Further out in the plasma, ETG modes are found to drive stiff transport, consistent both in magnitude and in flux ratios with predictions from interpretive transport modelling. These modes exhibit strong sensitivity to $\mathrm{E \times B}$ shear. In particular, outside the $q=3/2$ surface, enhanced $\mathrm{E \times B}$ shear suppresses ETG turbulence, leading to steeper temperature gradients in this region.

\section{H-mode discharge}
\label{sec:h_mode}

MAST-U achieves its highest performance in H-mode, making it essential to understand electron-scale transport in this regime. To this end, a high-performance H-mode discharge, \#48657, is analysed. This discharge reaches a peak $\beta_N = 4.0$ at $t = 0.53\,\mathrm{s}$, with a plasma current $I_\mathrm{p} = 750\,\mathrm{kA}$ and heating from both on-axis SS and off-axis SW neutral beams, giving a total injected power of $P_{\mathrm{NBI}} = 3.4\,\mathrm{MW}$.

\begin{figure}[!htb]
    \begin{subfigure}{0.49\textwidth}
        \centering
        \includegraphics[width=75mm]{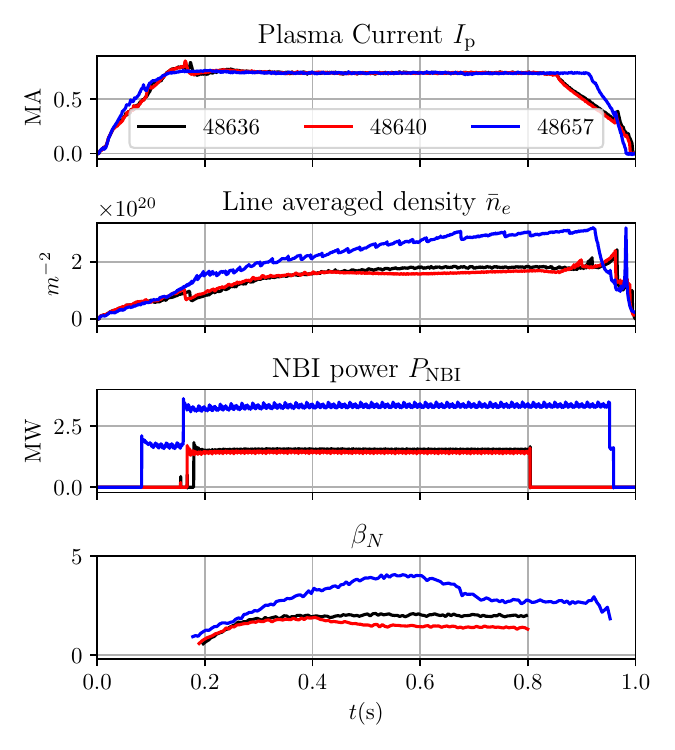}
        \caption{}
        \label{fig:hmode_pulse}
    \end{subfigure}
    \begin{subfigure}{0.49\textwidth}
        \centering
        \includegraphics[width=75mm]{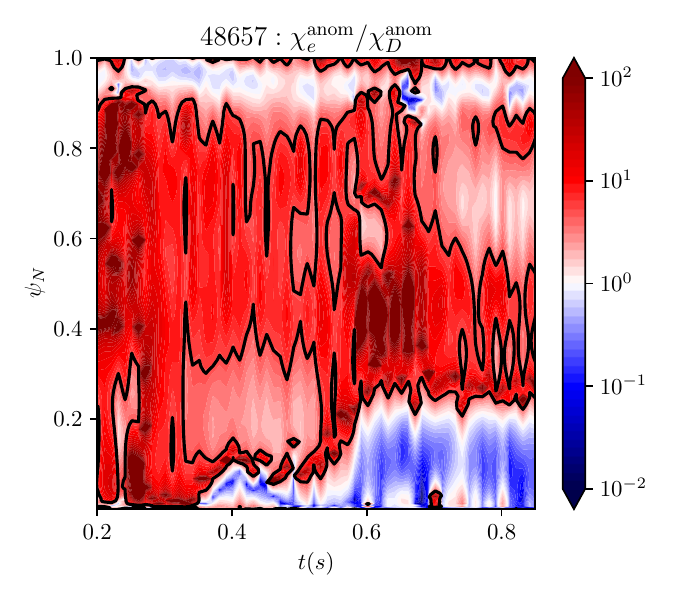}
        \caption{}
        \label{fig:48657_chi_ratio}
    \end{subfigure}
    \caption{(a) Time traces of plasma current, line-averaged density, NBI power, and $\beta_N$ for two L-mode MAST-U discharges (\#48636 in black and \#48640 in red) and the H-mode discharge \#48657 in blue. (b) Ratio of electron to deuterium heat diffusivity for \#48657 calculated using TRANSP, shown as a function of time and normalised poloidal flux. The black line denotes $\chi_e^{\mathrm{anom}} / \chi_D^{\mathrm{anom}} = 5$.}
    \label{fig:hmode_summary}
\end{figure}

Time traces of $I_\mathrm{p}$, line-averaged density $\bar{n}_e$, $P_{\mathrm{NBI}}$, and $\beta_N$ for \#48657 are shown in Figure \ref{fig:hmode_pulse}, alongside the two previously analysed L-mode discharges. As is typical for MAST-U H-mode plasmas, the density increases throughout the pulse due to strong particle confinement \cite{imada2024elm}. The full profiles of electron density, electron temperature and ion temperature are included for \#48657 in Figure \ref{fig:hmode_profiles} for completeness.

A rollover in $\beta_N$ and a rotation collapse are observed in \#48657, similar to the L-mode pulse \#48640. However, toroidal mode analysis indicates that the MHD activity is associated with an $m/n = 2/1$ mode and occurs later, at $t = 0.6\,\mathrm{s}$. The ratio of anomalous heat diffusivities, shown in Figure \ref{fig:48657_chi_ratio}, suggests a more complex transport picture. In the pedestal region, $\chi_e^{\mathrm{anom}} / \chi_D^{\mathrm{anom}} \sim 1$. Prior to the rotation collapse, during the high-$\beta_N$ phase, electron heat transport dominates over $0.4 \leq \psi_N \leq 0.8$, with $\chi_e^{\mathrm{anom}} / \chi_D^{\mathrm{anom}} \sim 10$. In the deep core, the transport across species becomes comparable, similar to the L-mode case. During the rotation collapse, a spike in the diffusivity ratio appears near $\psi_N = 0.4$, followed by a reduction around $\psi_N = 0.6$ in the final steady state.

\begin{figure}[!htb]
    \centering
    \includegraphics[width=75mm]{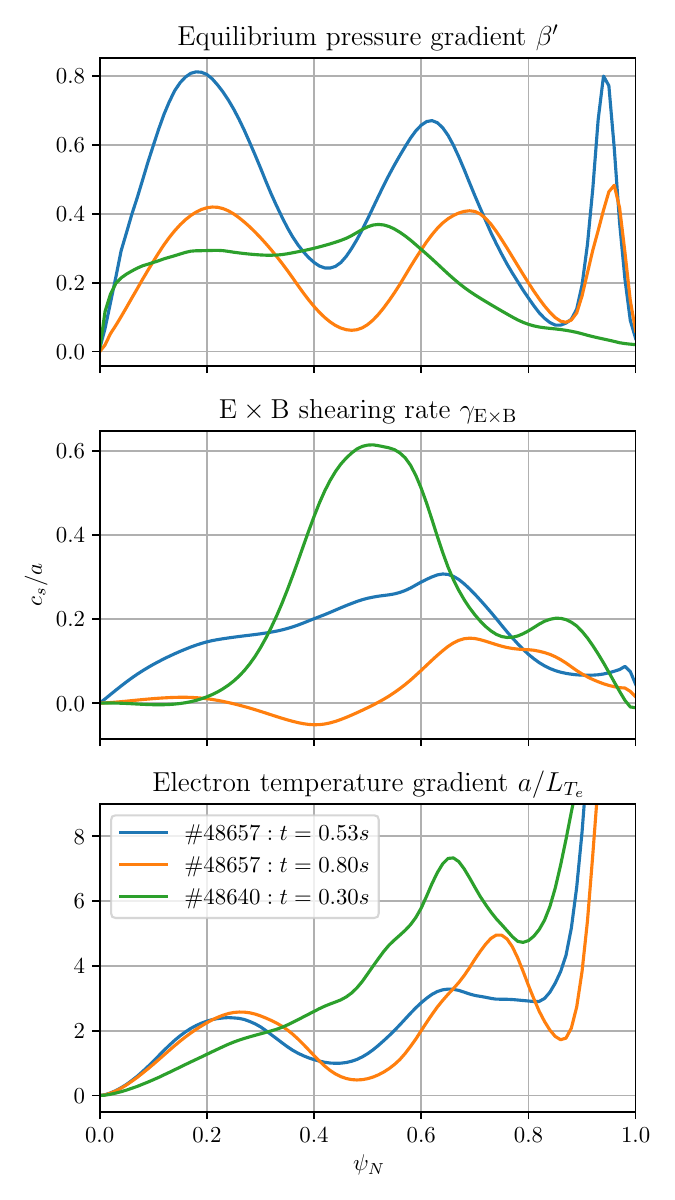}
    \caption{$|\beta'|$ (top), $\mathrm{E \times B}$ shearing rate (middle), and electron temperature gradient (bottom) for \#48657 at $t = 0.53\,\mathrm{s}$ (blue) and $t = 0.8\,\mathrm{s}$ (orange), compared with \#48640 at $t = 0.3\,\mathrm{s}$ (green).}
    \label{fig:48657_hmode_gradients}
\end{figure}

Several regions are identified where electron-scale transport may play a significant role. Figure \ref{fig:48657_hmode_gradients} shows profiles of $|\beta'|$, $\gamma_\mathrm{E \times B}$, and $a/L_{T_e}$ for \#48657 at peak $\beta_N$ ($t = 0.53\,\mathrm{s}$, blue) and after the rotation collapse ($t = 0.8\,\mathrm{s}$, orange). For reference, profiles from \#48640 before the rotation collapse are also shown (green). A local minimum in $|\beta'|$ for the H-mode case arises from a density flattening around $\psi_N = 0.4$. Across most of the core, $a/L_{T_e}$ is lower than in L-mode, while $|\beta'|$ is higher. Both effects act to stabilise ETG modes, suggesting reduced linear drive. Additionally, $\gamma_\mathrm{E \times B}$ is larger, providing further stabilisation. Following the rotation collapse, both $|\beta'|$ and $\gamma_\mathrm{E \times B}$ are significantly reduced throughout the inner core ($\psi_N \lesssim 0.6$), where $a/L_{T_e}$ nonetheless remains low. This reduction in stabilising mechanisms may allow ETG modes to drive more transport in the post-collapse phase. The outer core behaves differently. From $\psi_N = 0.65$ to $\psi_N = 0.8$, $a/L_{T_e}$ increases substantially, while $|\beta'|$ is largely unchanged or slightly increased; at $\psi_N = 0.8$, $\gamma_\mathrm{E \times B}$ is also comparable to its pre-collapse value. Any ETG destabilisation at these radii is therefore expected to arise from the steepening of the electron temperature profile rather than from a weakening of the stabilising terms.

\subsection{Linear stability in H-mode}

Linear simulations were performed at two times in the discharge: $t = 0.53\,\mathrm{s}$, corresponding to peak $\beta_N$, and $t = 0.8\,\mathrm{s}$, corresponding to the final steady state after the core rotation collapse. These simulations are fully electromagnetic, including both $A_{||}$ and $B_{||}$ fluctuations, and include a fast deuterium species modelled as a hot Maxwellian. The most relevant input parameters can be found in Tables \ref{tab:48657_t053} and \ref{tab:48657_t08}, with the full input files available in \ref{app:a}. Several core radii were examined, and in general ETG modes were found to be stable for $\psi_N \leq 0.4$.

\begin{figure}[!htb]
    \centering
    \includegraphics[width=150mm]{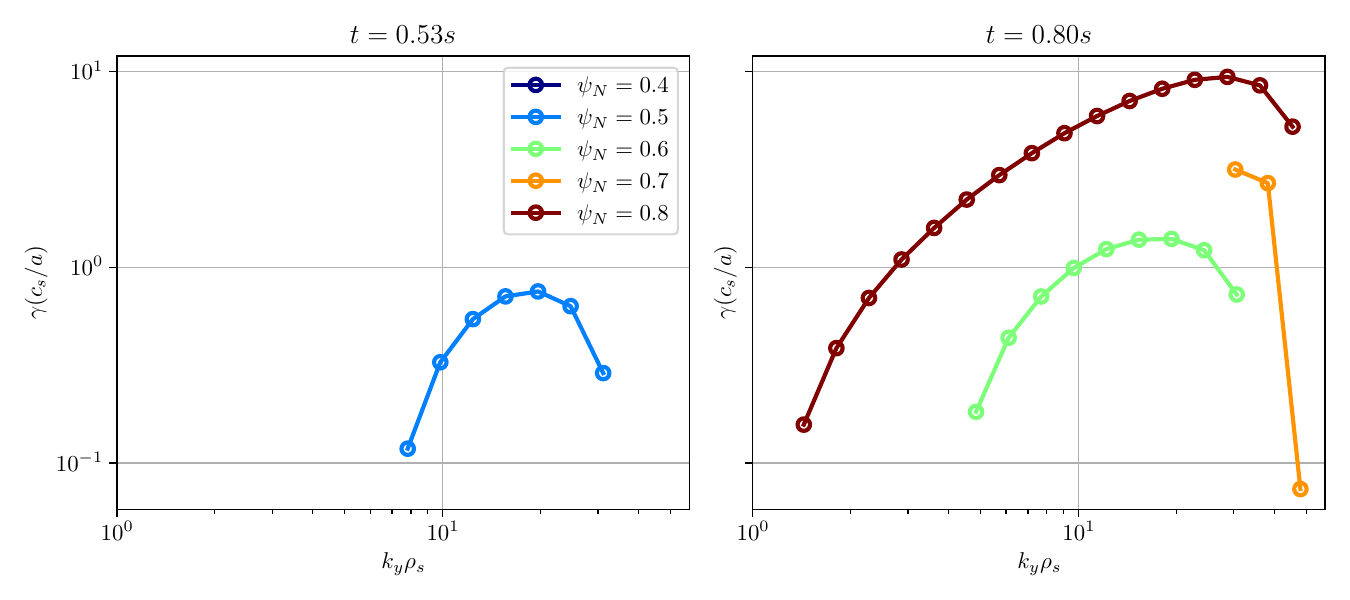}
    \caption{Growth rates of ETG modes in \#48657 at (left) $t = 0.53\,\mathrm{s}$ and (right) $t = 0.8\,\mathrm{s}$. Colours indicate the flux surface. Many surfaces are fully stable to ETG modes at both times. Note all the surfaces were simulated, so any surface that does not appear on the figure is completely stable at these scales.}
    \label{fig:48657_multi_surface}
\end{figure}

Figure \ref{fig:48657_multi_surface} shows the linear growth rates for modes with $k_y\rho_s > 1$. At peak $\beta_N$, only the surface at $\psi_N = 0.5$ exhibits unstable ETG modes, near a local minimum in $|\beta'|$ (see Figure \ref{fig:48657_hmode_gradients}). This stability at electron scales is consistent with previous studies in high-$\beta$ spherical tokamaks \cite{patel2022linear, kennedy2023electromagnetic}, and is primarily attributed to the relatively high $|\beta'|$ combined with lower $a/L_{T_e}$. The pedestal increases $\beta$ significantly, allowing for a stronger $\beta'$ stabilisation at a given pressure gradient, while also reducing the core gradient at fixed performance. Artificially setting $\beta'$ to the equivalent L-mode value destabilises the ETG across all wavenumbers, highlighting the strong sensitivity these ETG modes have to $\beta'$.

After the rotation collapse, surfaces in the inner core ($\psi_N < 0.6$) remain stable, while outer-core surfaces ($\psi_N \geq 0.6$) exhibit unstable ETG modes that may be experimentally relevant. Several ion-scale instabilities, such as KBMs and MTMs, are also present at $k_y\rho_s \leq 1$, but are not shown here.

\begin{figure}[!htb]
    \centering
    \includegraphics[width=150mm]{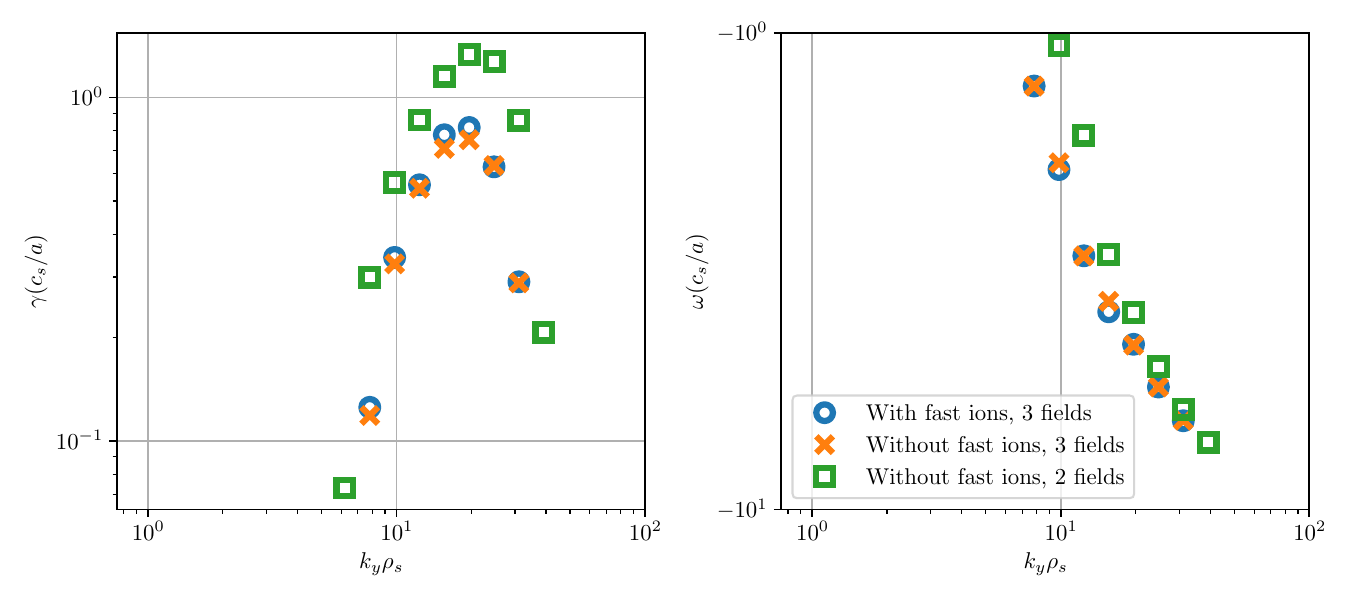}
    \caption{Linear $k_y$ spectrum for $\psi_N = 0.5$ showing simulations with all four kinetic species and three fields (blue circles), thermalised fast ions (orange crosses), and thermalised fast ions with no $B_{||}$ fluctuations (green squares).}
    \label{fig:48657_nofast}
\end{figure}

As in the L-mode case, the sensitivity of ETG modes to fast ions and $B_{||}$ fluctuations is examined in Figure \ref{fig:48657_nofast} for the $\psi_N = 0.5$ surface at $t = 0.53\,\mathrm{s}$. Removing fast ions has little impact on the growth rates, whereas removing $B_{||}$ fluctuations leads to a strong destabilisation, increasing the growth rate by nearly a factor of two, consistent with \cite{belli2010fully, roach2005microstability}. This demonstrates that inclusion of full electromagnetic effects is essential for nonlinear ETG simulations at high $\beta$. As before, fast ions are removed kinetically but their contribution to the equilibrium pressure gradient, $\beta'$, is retained when calculating drift frequencies.

\subsection{Nonlinear simulations in H-mode}

\begin{figure}[!htb]
    \centering
    \includegraphics[width=150mm]{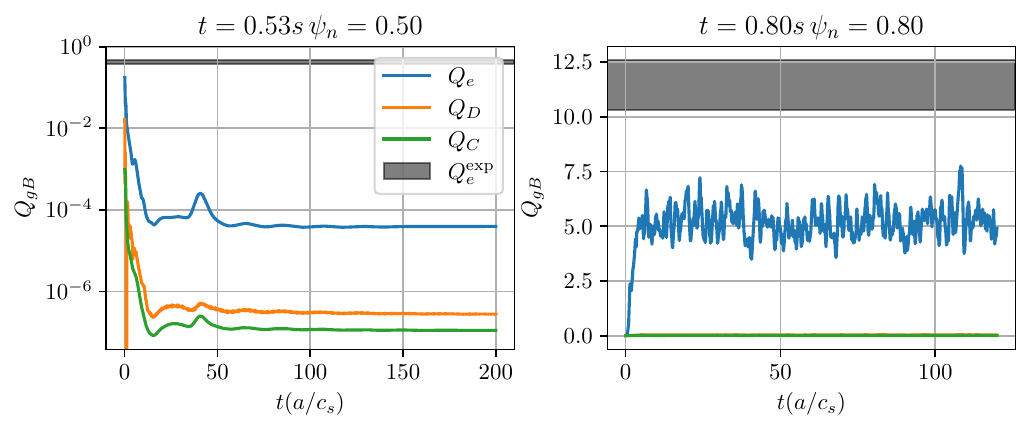}
    \caption{Nonlinear simulations of \#48657 showing the time evolution of the heat flux, separated by species. Left: $\psi_N = 0.5$ at $t = 0.53\,\mathrm{s}$. Right: $\psi_N = 0.8$ at $t = 0.8\,\mathrm{s}$. The grey bar indicates the experimental level of anomalous heat flux.}
    \label{fig:hmode_flux_t531}
\end{figure}

Nonlinear simulations were performed to assess the impact of the limited number of unstable ETG modes identified in the linear analysis. Prior to the rotation collapse, only the $\psi_N = 0.5$ surface exhibited unstable ETG modes, and this surface was therefore examined. Due to the narrow $k_y$ range over which ETGs are unstable, relatively low resolution was sufficient: 16 binormal modes with $k_{y,\min}\rho_s = 3.0$ and 256 radial modes with $k_{x,\min}\rho_s = 1.1$. Figure \ref{fig:hmode_flux_t531} shows the resulting heat flux, which is several orders of magnitude lower than the experimental value. This clearly indicates that ETG modes do not play a significant role in determining the confinement level in this phase.

After the rotation collapse at $t = 0.8\,\mathrm{s}$, the surface exhibiting the strongest ETG instability, $\psi_N = 0.8$, was analysed. At this location, $\chi_e^{\mathrm{anom}} / \chi_D^{\mathrm{anom}} > 5$, suggesting that ETG-driven transport may be relevant. To capture the full range of unstable modes and resolve radial streamers, higher resolution was required: 48 binormal modes with $k_{y,\min}\rho_s = 1.0$ and 384 radial modes with $k_{x,\min}\rho_s = 0.2$. The saturated heat flux, shown in Figure \ref{fig:hmode_flux_t531}, reaches approximately $50\%$ of the experimental value. Increasing the electron temperature gradient by 20\% brings the simulation into agreement with the experimental flux, although this lies outside the estimated uncertainty bounds. Nevertheless, the predicted transport is consistent with the diffusivity ratios obtained from TRANSP.

Overall, at the times analysed in this high-performance H-mode, ETG modes do not drive significant transport and are therefore unable to account for the levels inferred from interpretive modelling. This is due to various stabilising factors such as the high $|\beta'|$, relatively low electron temperature gradients, and strong rotation. Following the rotation collapse, ETG modes remain largely stable in the core, but can contribute appreciably to transport in the outer core. Direct comparison with experimental fluctuation measurements would further validate these simulations; current Doppler backscattering diagnostics \cite{shi2023first} probe up to intermediate scales ($k_y\rho_s \sim 4$), while proposed high-$k$ scattering systems \cite{spiers2026high} would enable access to smaller scales and help confirm the suppression of ETG modes at high $\beta$, which is critical for reactor-relevant conditions. 
We note that ETG turbulence has previously been found to be experimentally relevant close to mid-radius where the density gradient was low in the MAST H-mode pulse \#6252 \cite{applegate2004microstability, roach2005microstability, joiner2006electron}. High-$k$ scattering measurements of density fluctuations in the outer core of modest-$\beta$ NSTX H-mode discharges have also shown compelling agreement with gyrokinetic simulations of ETG turbulence \cite{ruiz2019validation}. Interest in the potential importance of ETG transport has been reignited by recent multi-scale simulations including both ion-scale MTMs and electron-scale ETGs \cite{belli2025separability}.

\section{Conclusion}
\label{sec:conclusion}

Electron-scale turbulent transport in MAST-U has been investigated using local gyrokinetic simulations with CGYRO, focusing on the role of electron temperature gradient modes in two L-mode plasmas and one H-mode plasma. Integrated modelling with TRANSP indicates that anomalous electron heat transport can dominate over ion transport under a range of conditions, motivating a detailed study of electron-scale instabilities.

In these L-mode discharges, ETG modes are found to be linearly unstable across much of the plasma outside $\psi_N\sim 0.4$, with growth rates peaking near $\psi_N=0.7$. The modes are predominantly toroidal in character, consistent with earlier microstability analyses of MAST and MAST-like equilibria \cite{applegate2004microstability, roach2005microstability, joiner2006electron}. Linear sensitivity scans show that ETGs are strongly destabilised by the electron temperature gradient $a/L_{T_e}$, stabilised by increasing $Z_\mathrm{eff}$ \cite{reshko2008effects} and $|\beta'|$ \cite{patel2022linear}, and relatively insensitive to collisionality. The dependence on the safety factor $q$ is non-monotonic and arises from competing effects of parallel streaming and the pressure-gradient contribution to the magnetic drift frequency. This mechanism becomes increasingly relevant in the higher-$\beta$ regime of spherical tokamaks and can limit the applicability of a simple $R/L_{T_e,\mathrm{crit}} \propto \hat{s}/q$ scaling \cite{jenko2001critical}, derived in the large-aspect-ratio, low-$\beta$ and weakly shaped limit, at finite $\beta'$. 

Nonlinear simulations show that ETG turbulence can generate substantial electron heat transport via radially extended streamers \cite{jenko2000electron, jenko2002prediction}, exhibiting stiffness and strong sensitivity to both $a/L_{T_e}$ and the $\mathrm{E \times B}$ shearing rate, in line with earlier nonlinear ETG studies at spherical tokamak parameters \cite{guttenfelder2011resolving, guttenfelder2013progress, belli2024flow}. In the outer core, the predicted heat flux agrees with experimental estimates within uncertainties, indicating that ETG modes can play a dominant role, as has been inferred from high-$k$ scattering measurements on NSTX \cite{ruiz2019validation}. In contrast, towards the inner core, particularly at high rotation, ETG-driven transport is reduced and cannot fully account for the observed anomalous heat flux, suggesting that additional mechanisms, such as ion-scale turbulence, are required. Following the core rotation collapse driven by the $m/n = 3/2$ neoclassical tearing mode, the reduction in $\gamma_\mathrm{E \times B}$ at $\psi_N = 0.5$ leads to a zonal-flow-dominated saturation state. In this regime, slowly growing zonal modes suppress ETG transport below experimental levels, consistent with the long-timescale zonal build-up identified by G. Colyer \textit{et al.} \cite{colyer2017collisionality}. This transition to a zonally dominated state is sensitive to the linear drive: modest increases in $a/L_{T_e}$ prevent strong zonal build-up and instead yield heat fluxes exceeding experimental values, qualitatively analogous to the drive threshold separating zonally dominated and strongly driven states in ion-scale ITG turbulence \cite{dimits2000comparisons, ivanov2020zonally, ivanov2022dimits}.

In the H-mode discharge \#48657, ETG modes are generally stable or only weakly unstable throughout the high-$\beta_N$ phase, and the associated nonlinear heat fluxes are orders of magnitude below experimental levels. This stability is primarily due to elevated $|\beta'|$ and lower $a/L_{T_e}$ compared to L-mode, consistent with previous observations in high-$\beta$ spherical tokamaks \cite{patel2022linear, kennedy2023electromagnetic}. At these higher $\beta$, parallel magnetic field fluctuations are no longer negligible: neglecting $B_{||}$ fluctuations nearly doubles the ETG growth rate, in line with fully electromagnetic eigenmode analyses of high-$\beta$ shaped plasmas \cite{belli2010fully, roach2005microstability}, so their inclusion is essential for quantitative electron-scale predictions in this regime. After the core rotation collapse, ETG modes become unstable towards the outer core ($\psi_N \geq 0.6$) and can drive transport approaching experimental levels at $\psi_N = 0.8$, where $a/L_{T_e}$ has increased. Overall, this analysis suggests that ETG turbulence is unlikely to be the dominant mechanism governing electron heat transport in high-performance H-mode plasmas \cite{kaye2021thermal}, motivating the examination of ion scales \cite{kennedy2026transition}.

In summary, ETG turbulence is most relevant in the outer core of MAST-U L-mode plasmas, where strong electron temperature gradients drive stiff electron heat transport, while the turbulence is stabilised by $\mathrm{E \times B}$ shear. The sensitivity to $Z_\mathrm{eff}$ \cite{reshko2008effects, romanelli2011linear}, which was not directly measured in these discharges, represents a key source of uncertainty and highlights the need for improved diagnostics. Ion-scale simulations, presented in a companion study \cite{kennedy2026tbd}, are required to explain transport in the inner core where ETG modes are stabilised. In addition, multi-scale interactions may become important in regimes where ion-scale turbulence is not fully suppressed and should be explored in future work \cite{belli2025separability, howard2014synergistic, hardman2019scale}.

\section*{Acknowledgements}
The authors would like to thank P. Ivanov and Y. Zhang for useful discussions regarding zonal saturation, and C. Beckley for discussions regarding uncertainties in experimental data. Simulations have been performed on the Leonardo National Supercomputing Consortium CINECA (Italy) under the project QLTURB. This work was performed using resources provided by the Cambridge Service for Data Driven Discovery (CSD3) operated by the University of Cambridge Research Computing Service (\url{www.csd3.cam.ac.uk}), provided by Dell EMC and Intel using Tier-2 funding from the Engineering and Physical Sciences Research Council (capital grant EP/T022159/1), and DiRAC funding from the Science and Technology Facilities Council (\url{www.dirac.ac.uk}). This work was supported by the Engineering and Physical Sciences Research Council [EP/R034737/1]. The work of T. Adkins was supported in part by the Laboratory Directed Research and Development (LDRD) Program at the Princeton Plasma Physics Laboratory for the U.S. Department of Energy under Contract No. DE-AC02-09CH11466. The United States Government retains a non-exclusive, paid-up, irrevocable, world-wide license to publish or reproduce the published form of this manuscript, or allow others to do so, for United States Government purposes.

\appendix

\section{Input parameters}
\label{app:a}

\begin{figure}[!htb]
    \begin{subfigure}{0.99\textwidth}
        \centering
        \includegraphics[width=150mm]{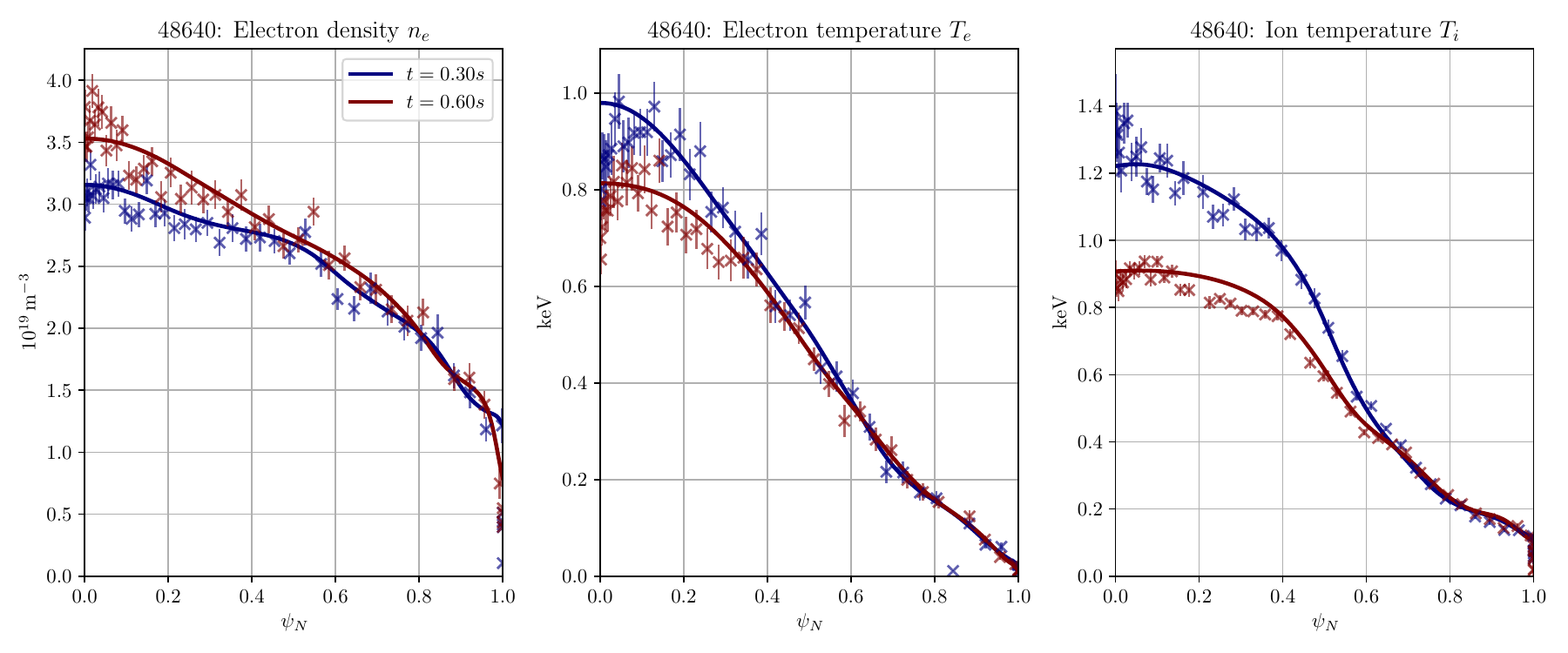}
        \caption{}
        \label{fig:lmode_profiles}
    \end{subfigure}
    \begin{subfigure}{0.99\textwidth}
        \centering
        \includegraphics[width=150mm]{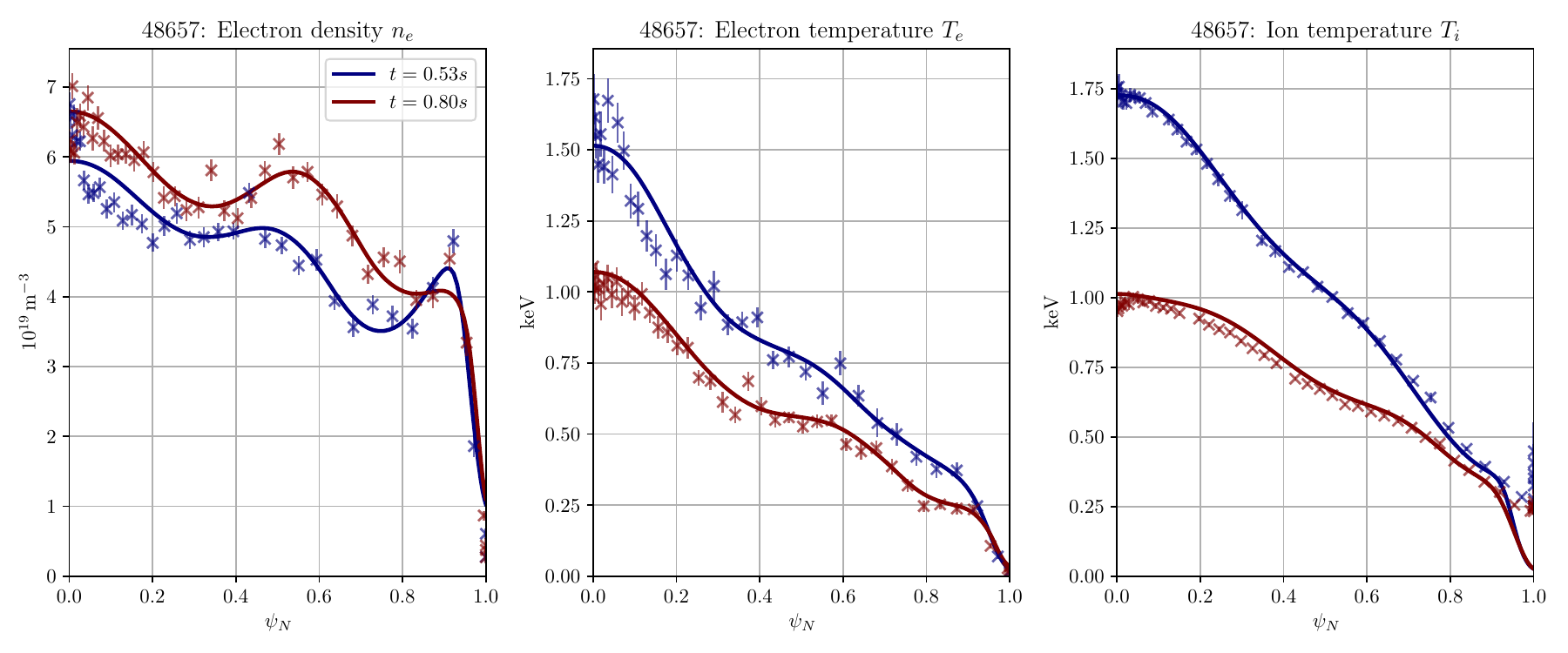}
        \caption{}
        \label{fig:hmode_profiles}
    \end{subfigure}
    \caption{Diagnostic data for the electron density, electron temperature and ion temperature along with the fits used in the TRANSP simulation for (a) \#48640 and (b) \#48657 before and after their corresponding rotation collapses.}
    \label{fig:profiles}
\end{figure}

Diagnostic measurements of the electron density, electron temperature and ion temperature from the Thomson scattering and charge exchange systems on MAST-U, along with the fits used in the TRANSP simulation, are shown in Figure \ref{fig:profiles} for both the L-mode and H-mode before and after the rotation collapse.

All inputs have been derived from TRANSP runs performed for these shots, with the corresponding TRANSP run IDs: 48636Q02, 48640Q02, 48657Q06. The input files were generated using Pyrokinetics at the specified flux surfaces and times. These baseline input files for CGYRO can be found at \url{https://github.com/bpatel2107/MASTU_ETG_2026_data_repo}. The most relevant input parameters are listed below for \#48640 and \#48657.

\begin{table}[htbp]
\centering
\begin{tabular}{l|ccccccc}
\hline
Parameter & \multicolumn{7}{c}{$\psi_N$} \\
\cline{2-8}
 & $0.3$ & $0.4$ & $0.5$ & $0.6$ & $0.7$ & $0.75$ & $0.8$ \\
\hline
$r/a$ & 0.468 & 0.546 & 0.617 & 0.685 & 0.754 & 0.789 & 0.826 \\
$R/a$ & 1.685 & 1.674 & 1.663 & 1.650 & 1.634 & 1.625 & 1.614 \\
$\kappa$ & 1.586 & 1.586 & 1.586 & 1.590 & 1.599 & 1.608 & 1.622 \\
$\delta$ & 0.037 & 0.040 & 0.044 & 0.049 & 0.057 & 0.064 & 0.074 \\
$q$ & 1.146 & 1.192 & 1.272 & 1.430 & 1.736 & 1.985 & 2.339 \\
$\hat{s}$ & 0.201 & 0.344 & 0.731 & 1.468 & 2.579 & 3.223 & 3.978 \\
$\beta_e$ & 0.027 & 0.023 & 0.019 & 0.012 & 0.007 & 0.005 & 0.004 \\
$\beta'$ & -0.280 & -0.300 & -0.362 & -0.296 & -0.165 & -0.118 & -0.079 \\
$a/L_{T_e}$ & 1.903 & 2.597 & 3.755 & 5.787 & 6.427 & 5.273 & 4.780 \\
$a/L_n$ & 0.375 & 0.318 & 0.809 & 1.756 & 1.445 & 1.404 & 1.836 \\
$\nu_{ee}\,(c_s/a)$ & 0.237 & 0.322 & 0.472 & 0.810 & 1.797 & 2.541 & 3.399 \\
$\gamma_{\mathrm{E \times B}}\,(c_s/a)$ & 0.132 & 0.444 & 0.613 & 0.510 & 0.210 & 0.158 & 0.172 \\
\hline
\end{tabular}
\caption{Flux surface parameters for the L-mode discharge \#48640 at $t = 0.3\,\mathrm{s}$.}
\label{tab:48640_t03}
\end{table}

\begin{table}[htbp]
\centering
\begin{tabular}{l|ccccccc}
\hline
Parameter & \multicolumn{7}{c}{$\psi_N$} \\
\cline{2-8}
 & $0.3$ & $0.4$ & $0.5$ & $0.6$ & $0.7$ & $0.75$ & $0.8$ \\
\hline
$r/a$ & 0.461 & 0.537 & 0.608 & 0.677 & 0.747 & 0.784 & 0.822 \\
$R/a$ & 1.680 & 1.672 & 1.662 & 1.651 & 1.637 & 1.628 & 1.618 \\
$\kappa$ & 1.520 & 1.521 & 1.525 & 1.537 & 1.558 & 1.574 & 1.595 \\
$\delta$ & 0.033 & 0.036 & 0.041 & 0.047 & 0.058 & 0.066 & 0.079 \\
$q$ & 1.031 & 1.080 & 1.188 & 1.389 & 1.740 & 2.006 & 2.370 \\
$\hat{s}$ & 0.154 & 0.498 & 1.100 & 1.890 & 2.755 & 3.275 & 3.820 \\
$\beta_e$ & 0.027 & 0.023 & 0.017 & 0.013 & 0.008 & 0.006 & 0.004 \\
$\beta'$ & -0.240 & -0.265 & -0.260 & -0.184 & -0.153 & -0.135 & -0.098 \\
$a/L_{T_e}$ & 1.618 & 2.718 & 3.780 & 4.291 & 5.832 & 6.066 & 5.414 \\
$a/L_n$ & 0.810 & 0.922 & 0.881 & 1.123 & 1.674 & 2.133 & 3.065 \\
$\nu_{ee}\,(c_s/a)$ & 0.300 & 0.383 & 0.565 & 0.900 & 1.663 & 2.375 & 3.290 \\
$\gamma_{\mathrm{E \times B}}\,(c_s/a)$ & 0.063 & 0.190 & -0.059 & -0.139 & 0.350 & 0.292 & 0.136 \\
\hline
\end{tabular}
\caption{Flux surface parameters for the L-mode discharge \#48640 at $t = 0.6\,\mathrm{s}$.}
\label{tab:48640_t06}
\end{table}

\begin{table}[htbp]
\centering
\begin{tabular}{l|ccccccc}
\hline
Parameter & \multicolumn{7}{c}{$\psi_N$} \\
\cline{2-8}
 & $0.3$ & $0.4$ & $0.5$ & $0.6$ & $0.7$ & $0.75$ & $0.8$ \\
\hline
$r/a$ & 0.458 & 0.539 & 0.615 & 0.689 & 0.761 & 0.797 & 0.834 \\
$R/a$ & 1.769 & 1.749 & 1.728 & 1.705 & 1.680 & 1.665 & 1.650 \\
$\kappa$ & 1.704 & 1.719 & 1.737 & 1.754 & 1.767 & 1.773 & 1.780 \\
$\delta$ & 0.100 & 0.114 & 0.131 & 0.151 & 0.175 & 0.190 & 0.207 \\
$q$ & 1.472 & 1.684 & 1.942 & 2.246 & 2.661 & 2.954 & 3.348 \\
$\hat{s}$ & 0.679 & 0.978 & 1.178 & 1.448 & 2.036 & 2.487 & 3.068 \\
$\beta_e$ & 0.072 & 0.064 & 0.059 & 0.044 & 0.028 & 0.024 & 0.022 \\
$\beta'$ & -0.533 & -0.259 & -0.385 & -0.657 & -0.452 & -0.283 & -0.155 \\
$a/L_{T_e}$ & 2.120 & 1.107 & 1.293 & 2.864 & 3.089 & 2.972 & 2.917 \\
$a/L_n$ & 0.330 & -0.392 & 0.608 & 2.660 & 1.750 & -0.070 & -1.798 \\
$\nu_{ee}\,(c_s/a)$ & 0.252 & 0.323 & 0.381 & 0.453 & 0.583 & 0.700 & 0.893 \\
$\gamma_{\mathrm{E \times B}}\,(c_s/a)$ & 0.165 & 0.201 & 0.249 & 0.288 & 0.259 & 0.185 & 0.116 \\
\hline
\end{tabular}
\caption{Flux surface parameters for the H-mode discharge \#48657 at $t = 0.53\,\mathrm{s}$.}
\label{tab:48657_t053}
\end{table}

\begin{table}[htbp]
\centering
\begin{tabular}{l|ccccccc}
\hline
Parameter & \multicolumn{7}{c}{$\psi_N$} \\
\cline{2-8}
 & $0.3$ & $0.4$ & $0.5$ & $0.6$ & $0.7$ & $0.75$ & $0.8$ \\
\hline
$r/a$ & 0.440 & 0.518 & 0.594 & 0.671 & 0.748 & 0.787 & 0.827 \\
$R/a$ & 1.716 & 1.706 & 1.693 & 1.678 & 1.658 & 1.646 & 1.632 \\
$\kappa$ & 1.498 & 1.520 & 1.555 & 1.603 & 1.657 & 1.683 & 1.707 \\
$\delta$ & 0.057 & 0.069 & 0.087 & 0.115 & 0.153 & 0.176 & 0.201 \\
$q$ & 1.128 & 1.323 & 1.638 & 2.072 & 2.608 & 2.923 & 3.300 \\
$\hat{s}$ & 0.689 & 1.303 & 1.796 & 2.048 & 2.168 & 2.322 & 2.630 \\
$\beta_e$ & 0.053 & 0.046 & 0.047 & 0.042 & 0.027 & 0.020 & 0.016 \\
$\beta'$ & -0.331 & -0.131 & -0.078 & -0.296 & -0.407 & -0.325 & -0.199 \\
$a/L_{T_e}$ & 2.458 & 1.233 & 0.520 & 2.004 & 4.204 & 4.948 & 3.397 \\
$a/L_n$ & 0.561 & -0.718 & -0.612 & 1.320 & 2.736 & 1.908 & 0.755 \\
$\nu_{ee}\,(c_s/a)$ & 0.511 & 0.692 & 0.814 & 0.930 & 1.259 & 1.640 & 2.174 \\
$\gamma_{\mathrm{E \times B}}\,(c_s/a)$ & -0.019 & -0.051 & -0.011 & 0.078 & 0.154 & 0.135 & 0.127 \\
\hline
\end{tabular}
\caption{Flux surface parameters for the H-mode discharge \#48657 at $t = 0.8\,\mathrm{s}$.}
\label{tab:48657_t08}
\end{table}

\clearpage
\printbibliography[]

\end{document}